\documentclass[AMA,STIX2COL]{WileyNJD-v2}
\usepackage{moreverb}

\articletype{}% Biotechnology \& Bioengineering

\received{<day> <Month>, <year>}
\revised{<day> <Month>, <year>}
\accepted{<day> <Month>, <year>}

\usepackage{colortbl}
\usepackage{xcolor}
\usepackage{algpseudocode}
\usepackage{bm}
\usepackage{amsmath}

\usepackage{mathtools}
\usepackage{subfig}
\usepackage{float}
\usepackage{multicol}
\usepackage{siunitx}
\usepackage{comment}

\begin{document}
\title{Metabolic Regulatory Network Kinetic Modeling with Multiple Isotopic Tracers for iPSCs}

\author[1]{Keqi Wang}
\author[1]{Wei Xie  $^{*,}$}
\author[2]{Sarah W. Harcum $^{\dagger,}$}

\authormark{Wang \textsc{et al}}

% Include full affiliation details for all authors
\address[1]{Department of Mechanical and Industrial Engineering, Northeastern University, Boston, MA 02115, USA}
\address[2]{Department of Bioengineering, Clemson University, Clemson, SC, USA}

\corres{$^{*}$Department of Mechanical and Industrial Engineering, Northeastern University, Boston, MA 02115, USA. \\
\email{w.xie@northeastern.edu}\\
$^{\dagger}$Department of Bioengineering, Clemson University, Clemson, SC, USA. \\
\email{harcum@clemson.edu}}

% Here goes the abstract
\abstract[Abstract]{
The rapidly expanding market for regenerative medicines and cell therapies highlights the need to advance the understanding of cellular metabolisms and improve the prediction of cultivation production process for human induced pluripotent stem cells (iPSCs). In this paper, a metabolic kinetic model was developed to characterize the underlying mechanisms of iPSC culture process, 
which can predict cell response to environmental perturbation and support process control. This model focuses on the central carbon metabolic network, including glycolysis, pentose phosphate pathway (PPP), tricarboxylic acid (TCA) cycle, and amino acid metabolism, which plays a crucial role to support iPSC proliferation. Heterogeneous measures of extracellular metabolites and multiple isotopic tracers collected under multiple conditions were used to learn metabolic regulatory mechanisms. Systematic cross-validation confirmed the model's performance in terms of providing reliable predictions on cellular metabolism and culture process dynamics under various culture conditions. Thus, the developed mechanistic kinetic model can support process control strategies to strategically select optimal cell culture conditions at different times, ensure cell product functionality, and facilitate large-scale manufacturing of regenerative medicines and cell therapies.
%\sep Partially observable markov decision process (MDP)
%Stochastic decision process 
%\sep
%Stochastic process optimization %Optimal Design of Experiments, Digital Twin
}


\maketitle

%\clearpage
\section{Introduction}
\label{sec:introduction}  
Cell therapeutics and regenerative medicines have the potential to treat and prevent diseases, such as cancers, and cardiovascular and hematologic diseases \citep{hanna2016advanced,stephanopoulos1998metabolic}. The cell therapy market is experiencing unparalleled growth and the projected worth of the relevant market is expected to be over \$8 billion in 2025 \citep{fiorenza2020value}. To meet the rising demand, cost reductions in manufacturing and increases in quality for human induced pluripotent stem cells (iPSCs) on a large scale are crucial for the success of cell therapies \citep{odenwelder2021induced}.  Unlike traditional biopharmaceuticals, the cells are the product with functional identity depending on the regulatory metabolic dynamic behaviors. 

The efficacy of the cells is very sensitive to culture conditions. Variability in the culture conditions can lead to reduced yields and heterogeneous cell populations.  Heterogeneous cell populations have been observed to increase the potential for tumor or teratoma formulation in the patient \citep{dressel2011effects}. Traditional cell culture process control strategies often are ad hoc and rely solely on experimental approaches or PID controllers ignoring long-term effects.  The optimal design and control of mammalian cell culture processes, 
especially iPSCs, can be laborious due to the limited understanding of cellular metabolisms and end-to-end cultivation process dynamics \citep{kyriakopoulos2018kinetic,wang2021cell}.

\begin{sloppypar}
In order to determine the best culture conditions, accounting for cell life cycle, relationships between culture conditions and the output trajectory (i.e., cell functional identity and yield) are needed. {Mechanistic dynamics model has the capability to predict cell outcomes (e.g., metabolic flux rates) and can be used to 
assess product quality
%. These models provide systems-level descriptions of metabolic networks and regulatory mechanisms.  With this complex and high-level cellular description, it is possible to interface such models with process control algorithms to 
and strategically select the optimal culture conditions at different times.}
\end{sloppypar}

%The proposed model for the iPSC metabolic network and regulation mechanisms combines four modeling approaches to leverage the strengths of each approach: (1) metabolic flux analysis (MFA), (2) flux balance analysis (FBA), (3) $^{13}$C-MFA, and (4) kinetic models. 
{%Drawing inspiration from three well-established modeling approaches, 
The proposed mechanistic model describes an iPSC metabolic network regulatory mechanisms. %leverages the strengths of each: 
It is associated with the existing studies, including
(1) metabolic flux analysis (MFA) \citep{niklas2012metabolic,nolan2011dynamic}, (2) $^{13}$C-MFA \citep{antoniewicz2015methods,antoniewicz2018guide,rivera2021critical,sengupta2011metabolic,wiechert2001universal,quek2010metabolic,dong2019dissecting,leighty2011dynamic}, and (3) kinetic models \citep{kyriakopoulos2018kinetic,nolan2011dynamic,ghorbaniaghdam2014analyzing,ghorbaniaghdam2013kinetic,ghorbaniaghdam2014silico}.}
In MFA, metabolic flux rates (such as substrate uptake rate and metabolite secretion rate) are estimated based on experimental measurements subject to stoichiometric constraints. Under the standard assumption of (pseudo) steady state for intracellular metabolites, the sum of all fluxes producing a metabolite is equal to the sum of all fluxes consuming that metabolite. This technique is frequently used to compare the metabolism of different cell lines to assess the activity of individual pathways under different cultivation conditions \citep{niklas2012metabolic}. However, intracellular metabolites can change during cell proliferation \citep{templeton2013peak}.

Stable isotope studies, integrating with MFA, have provided more detailed information on intracellular state and metabolic pathways, namely $^{13}$C-MFA can precisely estimate metabolic reaction rates.  {Typically these flux rates are estimated from measured intracellular mass isotopomer distribution %isotopic distributions 
(MID)  patterns and external metabolites concentrations. 
Common $^{13}$C and $^{2}$H isotopic tracers were utilized for investigating mammalian cell metabolism with a specific emphasis on the interpretation of isotopic labeling patterns\cite{dong2019dissecting}.
Many existing studies are built on the assumption of isotopic and metabolic steady state \citep{antoniewicz2015methods,antoniewicz2018guide,rivera2021critical,sengupta2011metabolic,wiechert2001universal,quek2010metabolic,dong2019dissecting}. Several modeling tools have been developed for situations involving isotopic and/or metabolic non-steady states \citep{leighty2011dynamic,antoniewicz2015methods, antoniewicz2018guide}.}
%However, these tools are not easily integrated into the process control of iPSC cultures.
 % {(rewrite this sentence and need ref here)}. 
\begin{sloppypar}

Another approach to model cell dynamics includes kinetic-metabolic models \citep{kyriakopoulos2018kinetic}. The vast majority of the kinetic models use Monod and Michaelis–Menten expression formalisms modeling the metabolic regulation mechanisms 
 \citep{ghorbaniaghdam2014analyzing,ghorbaniaghdam2013kinetic,ghorbaniaghdam2014silico,kyriakopoulos2018kinetic,nolan2011dynamic}.  For Chinese hamster ovary (CHO) cell cultures,  kinetic-metabolic models include glycolysis, pentose phosphate pathway (PPP), tricarboxylic acid (TCA) cycle, respiratory chain, as well as the regulatory functions from energy shuttles (ATP/ADP) and cofactors (NADH, NAD$^+$, NADPH, NADP$^+$);  this high level of details is required to predict hypoxic perturbation \citep{ghorbaniaghdam2013kinetic,ghorbaniaghdam2014silico}. For example, the study \cite{nolan2011dynamic} proposed a Monod model characterizing the cell flux rate response to environmental perturbation. That study used limiting substrate kinetics to calculate the extracellular metabolite consumption/production rates and further infer metabolic flux rates built on a static MFA assumption, i.e., assuming the intracellular metabolites are at the pseudo-steady state. Even though several \textit{in silico} metabolomic platforms have been developed for CHO cells \citep{ghorbaniaghdam2014analyzing,ghorbaniaghdam2013kinetic,ghorbaniaghdam2014silico,kyriakopoulos2018kinetic,nolan2011dynamic}, a  metabolic regulatory network simulator for iPSC has not been developed. 
\end{sloppypar}

{As lactate accumulation emerges as an inevitable outcome of iPSC metabolism, it becomes essential to explore the consequences of accumulated extracellular lactate on intracellular iPSC metabolism. Equally important is to comprehend how glucose concentration acts as a regulatory factor, influencing lactate production. Understanding the implications of these concentration variations is not only critical for the survival and proliferation of iPSCs \citep{nelson2008lehninger} but also essential for maintaining the quality of cell products. Hence, the main objective of this study is to develop and validate a mechanistic model of an iPSC metabolic regulatory network that can precisely predict cell responses to environmental changes, with a particular focus on variations in extracellular glucose and lactate concentrations.} %This paper utilizes data collected from well-structured monolayer K3 iPSC cultures, as described in \cite{odenwelder2021induced}. %To support the prediction of iPSC responses to environmental changes, especially with regard to fluctuations in extracellular glucose and lactate concentrations during the culture process, 

{
The proposed model primarily focuses on central metabolism, given its pivotal role in stem cell proliferation, biosynthesis, and overall functionality.
%As lactate accumulation emerges as an inevitable outcome of iPSC metabolism, it becomes essential to explore the consequences of accumulated extracellular lactate on intracellular iPSC metabolism. Equally important is to comprehend how glucose concentration acts as a regulatory factor, influencing lactate production. Understanding the implications of these concentration variations is not only critical for the survival and proliferation of iPSCs \citep{nelson2008lehninger} but also essential for maintaining the quality of cell products. 
%Comprehending how extracellular lactate accumulation influences intracellular iPSC metabolism, in conjunction with the role of glucose concentration in regulating lactate production, is of paramount importance. T
%his understanding is pivotal for iPSC survival, proliferation, and the maintenance of high-quality cell products. Hence, the main objective of this study is to develop and validate a mechanistic model of an iPSC metabolic regulatory network that can precisely predict cell responses to environmental changes, with a particular focus on variations in extracellular glucose and lactate concentrations. 
%This paper utilizes data collected from well-structured monolayer K3 iPSC cultures, as described in \cite{odenwelder2021induced}.
%The developed mechanistic model characterizes time-varying reaction dynamics and regulatory mechanisms. 
To estimate model parameters, this model utilized experimentally obtained data from well-designed monolayer K3 iPSC cultures, as described in the study \cite{odenwelder2021induced}.
%for the time-variate reactions and regulatory mechanisms.
During the model development, the level of detail required for key pathways was explored, namely the PPP and branched amino acids uptake rates into the TCA cycle.  In this paper, the proposed kinetic model structure with critical regulatory mechanisms will be discussed and validated through comparisons with the experimental data, with the aim of ensuring the model's ability to accurately predict cell responses to environmental changes.
%The model development, validation, and final model details will be presented with comparisons to the experimental data, aiming to ensure the proposed mechanistic model can accurately predict cell responses to environmental changes.
}

The proposed mechanistic model, comprising the metabolic reaction network structure and flux regulatory mechanism, can be easily customized to suit various mammalian cell culture systems, e.g., embryonic stem cells (ESCs), CHO, etc. {This adaptability arises from the underlying principles and general characteristics of metabolic regulation that are common across various mammalian cells.} 
Through the collection of extracellular metabolite concentrations and intracellular MID measurements, this mechanistic model can be adeptly employed to analyze and forecast the behavior of other mammalian cell cultures with consistent accuracy and reliability.

% ===============================

\section{Data Description}
\label{sec:datadescription}

\begin{sloppypar}
Briefly, 
the data used to estimate the model parameters were derived from an experimental study of K3 iPSCs, where monolayer cultures were conducted in 6-well plates and 60 mm cell culture dishes\citep{odenwelder2021induced}. The data encompassed extracellular metabolite measurements and isotopic labeling details for four distinct culture conditions, utilizing three tracers: glucose, glutamine, and lactate .
%the data used to obtain model parameters were from an experimental study of K3 iPSCs that included extracellular metabolite and isotopic labeling information for four culture conditions using three tracers: glucose, glutamine, and lactate  \citep{odenwelder2021induced}. %comparisons.  
The four different initial culture media concentrations were: (1) high glucose and low lactate (HGLL); standard initial conditions, i.e., 18.3 mM glucose and 0 mM lactate; (2) high glucose and high lactate (HGHL), i.e., 18.3 mM glucose and 20 mM lactate; (3) low glucose and low lactate (LGLL), i.e., 5.6 mM glucose and 0 mM lactate; and (4) low glucose and high lactate (LGHL), i.e., 5.6 mM glucose and 20 mM lactate. Table~\ref{tab:media} presents the media concentrations used for each condition, encompassing glucose, lactate, and supplemented sodium chloride concentrations. Furthermore, 2.75 mM glutamine was included in each condition. These concentrations were chosen to ensure comparable growth rates across all conditions while maintaining pluripotency \citep{odenwelder2021induced}.

\begin{figure}[h!]
    \centering
    \includegraphics[width=0.5\textwidth]{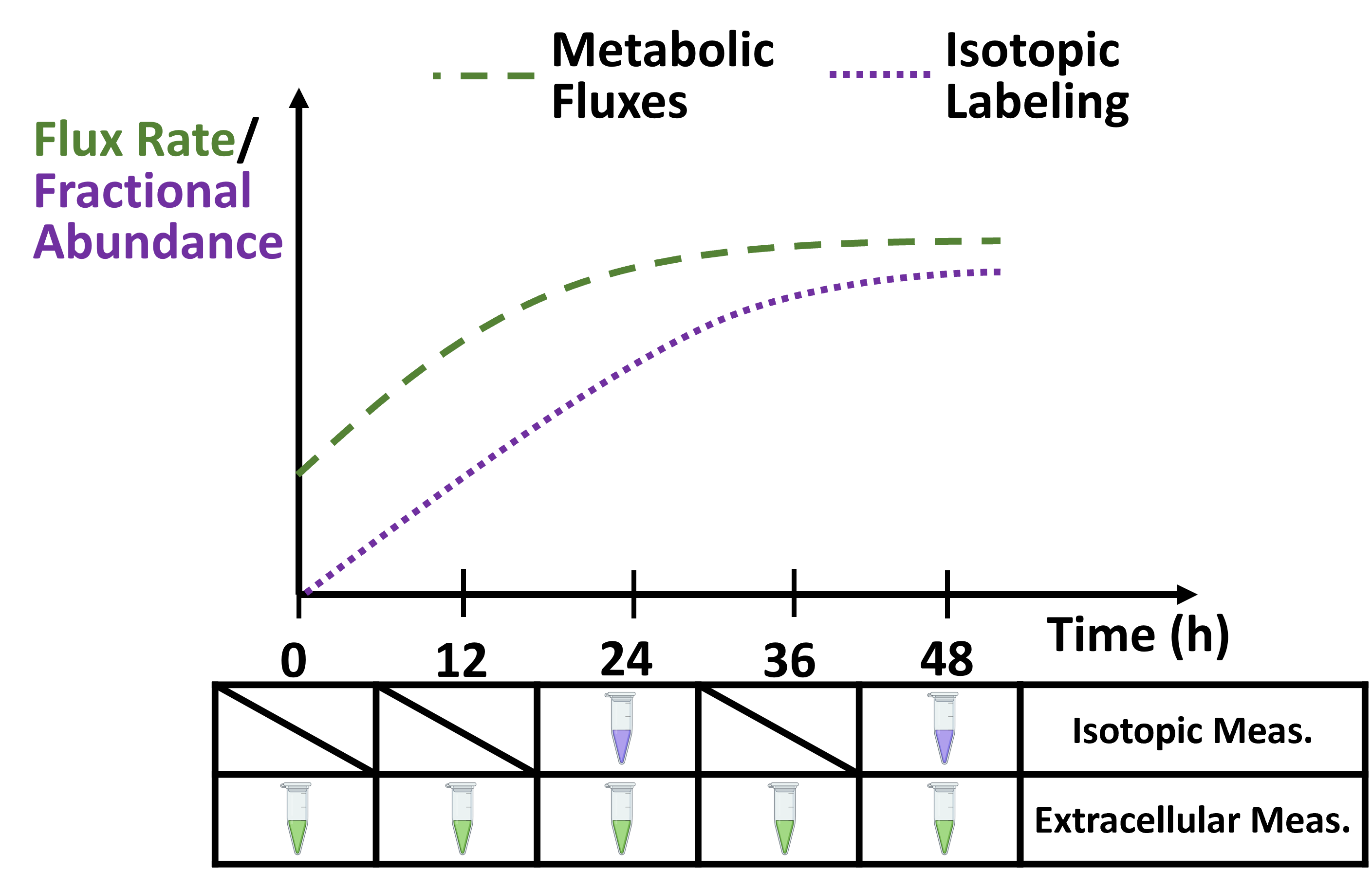}
    \caption{
    {
    Schematic for cell metabolic and isotopic states 
    when time-course measurements of extracellular concentrations and intracellular %amino acid 
    MIDs were taken during the experimental iPSC cultures 
    (Adapted from {Jazmin and Young} \cite{jazmin2013isotopically} and created with BioRender.com). Metabolic steady state is reached prior to isotopic steady state \cite{odenwelder2021induced}.
    %Metabolic concentration measurements were conducted every 12 hours (green Eppendorf tube icon), while the MIDs were acquired at 24-h and 48-h time points (purple Eppendorf tube icon). The 24-h measurements were considered to represent metabolic steady state, but not isotopic steady state, while by 48-h, both metabolic and isotopic steady states were achieved.
    %A schematic of the time-course extracellular concentrations measurements and isotopic intracellular amino acid MIDs for the experimental  iPSC cultures. Metabolics were measured every 12 h, while the MIDs were only obtained at 24- and 48-h. The 24-h MIDs were considered to be at metabolic steady-state, but not isotopic steady-state, while by 48-h, both metabolic and isotopic steady-state were reached. adopted from \cite{jazmin2013isotopically}
    } 
    }
    \label{fig:doe}
\end{figure}

{Cell density, glucose, lactate, pyruvate, and extracellular amino acid concentrations were obtained for each condition at 0-, 12-, 24-, 36-, and 48-h. Intracellular amino acid MIDs were obtained from parallel labeling experiments that used [1,2-$^{13}$C$_2$] glucose, [U-$^{13}$C$_5$] L-glutamine, and [U-$^{13}$C$_3$] sodium L-lactate (when lactate was added to the culture media). The time-course measurements, as illustrated in Figure~\ref{fig:doe}, included the transition %from a non-steady state 
to metabolic and isotopic steady states \cite{odenwelder2021induced}. These measurements were used to develop the metabolic regulatory network kinetic model for iPSC cultures.}
%Cell density, glucose, lactate, pyruvate, and extracellular amino acid concentrations were obtained for each condition at 0-, 12-, 24-, 36-, and 48-h.  Intracellular amino acid MIDs are obtained from parallel labeling experiments that used [1,2-$^{13}$C$_2$] glucose, [U-$^{13}$C$_5$] L-glutamine, and [U-$^{13}$C$_3$] sodium L-lactate (when lactate was added to culture media). A schematic of the time-course measurements as shown in Figure~\ref{fig:doe} were used to develop the metabolic regulatory network kinetic model for iPSC culture process. 

\begin{table}[h!]
\centering
\caption{The media formulations for K3 iPSC cultures include initial concentrations of glucose, sodium L-lactate, and supplemented sodium chloride (NaCl). NaCl was added to balance osmolarity in cultures lacking sodium lactate. Additionally, the media contains 2.75 mM L-glutamine.}
\label{tab:media}
\begin{tabular}{|c|cccc|}
\hline
\rowcolor[HTML]{DDEBF7} 
& \multicolumn{4}{c|}{\cellcolor[HTML]{DDEBF7}\textbf{Concentrations (mM)}}     \\ \hline
\rowcolor[HTML]{DDEBF7} 
\textbf{Component}                                     & \multicolumn{1}{c|}{\cellcolor[HTML]{DDEBF7}\textbf{HGLL}} & \multicolumn{1}{c|}{\cellcolor[HTML]{DDEBF7}\textbf{HGHL}} & \multicolumn{1}{c|}{\cellcolor[HTML]{DDEBF7}\textbf{LGLL}} & \multicolumn{1}{c|}{\cellcolor[HTML]{DDEBF7}\textbf{LGHL}} \\ \hline
\cellcolor[HTML]{DDEBF7}\textbf{Glucose}                & \multicolumn{1}{c|}{18.3}                               & \multicolumn{1}{c|}{18.3}                                & \multicolumn{1}{c|}{5.6}                       & \multicolumn{1}{c|}{5.6}          \\ \hline
\cellcolor[HTML]{DDEBF7}\textbf{Lactate}                & \multicolumn{1}{c|}{-}                                & \multicolumn{1}{c|}{20.0}                                & \multicolumn{1}{c|}{-}                       & \multicolumn{1}{c|}{20.0}           \\ \hline
\cellcolor[HTML]{DDEBF7}\textbf{Added NaCl}               & \multicolumn{1}{c|}{+20.0}                                 & \multicolumn{1}{c|}{-}                                 & \multicolumn{1}{c|}{+20.0}                       & \multicolumn{1}{c|}{-}               \\ \hline
\end{tabular}
\end{table}

\end{sloppypar}

% >>>>>>>

\section{Model Development}
\label{sec: Model}

%In this section, %we present the model development %of iPSC metabolic regulatory network kinetic model development 
%according to the real-world extra- and intra-cellular measurements. 
{
%The K3 iPSCs were cultured in a 2D monolayer on 6-well plates and 60 mm cell culture dishes, establishing a relatively homogeneous microenvironment. Consequently, this study disregards cellular heterogeneity and concentrates on characterizing the mean metabolic flux rate response for a uniform cell population. 
iPSCs cultured in petri-dish 2D monolayer have homogeneous environmental condition. Thus, the objective of this study is to develop a deterministic metabolic regulatory network kinetic model characterizing the mean metabolic dynamics, 
%This objective is accomplished using a deterministic metabolic regulatory network kinetic model, 
which is presented in four developmental stages. 
Section~\ref{subsec:kinetic_model} provides the general dynamic model description, including (1) metabolic reaction network kinetic model; and (2) %regulatory mechanism modeling and learning; and (3) 
time-course isotopic labeling simulation.} Section~\ref{subsec:reactionnetwork} presents the iPSC metabolic reaction network which includes central carbon metabolism. In Section~\ref{subsec:regulatorymechanism}, {the metabolic flux kinetic model}, characterizing cell response to environmental variations, is fitted by using time-course extracellular concentration measurements and MIDs. 
%the alternative biologically relevant mathematical representations of metabolic flux kinetics are presented. 
Finally, in Section~\ref{subsec: goodness-of-fit}, the assessment of  fit and validation of the dynamic model for the iPSC cultures is discussed, where the objective function %, model calibration, 
and goodness-of-fit will be presented.  

\subsection{Metabolic Network Kinetic Modeling}
\label{subsec:kinetic_model}
%{The general regulatory and metabolic network for the kinetic model as well as the Time-course Isotopic Labeling Simulation… …. need a transition or introduction.  the (1) and (2) are too abrupt. … A paragraph needs 5 sentence.  Intro, three details, summary}

In this section, the general regulatory metabolic network kinetic model and time-course isotopic labeling simulation are presented. The kinetic model captures the dynamic changes in cell density, and extracellular and intracellular metabolite concentrations, where the changing rates of metabolites depend on the concentration of substrates and inhibitors. To incorporate the mass isotopic data, a dynamic isotopic labeling system was constructed, and a time-course isotopic labeling simulation was developed. Overall, the metabolic reaction network kinetic model provides a comprehensive understanding of the underlying mechanisms of the iPSC cultures. %process.

\vspace{0.05in}

\textbf{(1) Metabolic Reaction Network Kinetic Model.} The data from the study for K3 iPSCs were collected during the exponential growth phase \citep{odenwelder2021induced}. %Growth rates were calculated from 0 to 36 hr and were similar between the four conditions (p > 0.05). The similar growth
%rates shown in Figure 2b confirm that the glucose and lactate concentrations
%did not significantly affect the K3 iPSCs growth rates. 
%By experimental design and the statistical analysis, the specific cell growth rates, denoted by $\mu(\mbox{h}^{-1})$, were determined to be constant for the four batch culture conditions, i.e., HGLL, HGHL, LGLL, and LGHL. i.e., that the glucose and lactate concentrations did not significantly affect the K3 iPSCs growth rates. 
Through experimental design and statistical analysis, our previous study \cite{odenwelder2021induced} discovered uniform specific cell growth rates ($\mu$) across the four batch culture conditions (HGLL, HGHL, LGLL, and LGHL), maintaining pluripotency. This suggests that glucose and lactate concentrations within these ranges did not notably influence K3 iPSC growth rates and did not have a detrimental impact on human K3 iPSC pluripotency.
%However, {the future conditions the growth rate may not be constrained,} 
Thus the classical cell density formalism was used to relate growth and  the cell density, denoted by $X(t)$ (cells/cm$^2$), at time $t$ as, 
\begin{equation}
    %X^\prime(t) 
    \frac{d X(t)}{dt} = \mu X(t).   
    \label{eq:celldensity}
\end{equation}

\begin{sloppy}   
The extracellular metabolite concentrations at time $t$ are denoted by a vector with dimension $p$, i.e., $\pmb{s}(t) = \big(s_1(t),s_2(t),\dots,s_p(t)\big)^\top$. Similarly, the intracellular metabolite concentrations of interest are denoted by a vector with dimension $q$ at time $t$ as  $\pmb{\ell}(t) = \big(\ell_1(t),\ell_2(t),\dots,\ell_q(t)\big)^\top$. 
Thus, at any time $t$, the state
of extracellular and intracellular metabolite concentrations is denoted by 
\begin{equation}
    \pmb{u}(t)=\big(\pmb{s}(t)^\top,\pmb{\ell}(t)^\top\big)^\top. 
    \nonumber 
\end{equation}
\end{sloppy}

%To model the dynamic evolution of cell response to environmental perturbation during the iPSC culture process, 
{To model the dynamic evolution of cell response to environmental fluctuations during the iPSC cultures,  at any time $t$, the specific reaction flux rates, represented by a vector with dimension $n$, depend on both extracellular and intracellular metabolite concentrations,  i.e.,} %}at any time $t$, the specific reaction flux rates represented by 
%a vector of with dimension $n$ depend on the extracellular and intracellular metabolite concentrations, i.e.,   
\begin{equation}
    \pmb{v}[\pmb{u}(t)] = \big(v_1[\pmb{u}(t)],v_2[\pmb{u}(t)],\dots,v_n[\pmb{u}(t)]\big)^\top.
    \nonumber
\end{equation}
Let $N$ denote a $(p+q) \times n$ stoichiometry matrix characterizing the structure of metabolic reaction network. 
Therefore, the dynamic evolution of extracellular and intracellular metabolite concentrations
is modeled through a mass balance on the system of equations,
\begin{equation}
    %\pmb{u}^\prime(t) 
    \frac{d\pmb{u}(t) }{dt}
    = N  \pmb{v}[\pmb{u}(t)]  X(t).   
    \label{eq:state}
\end{equation}

\vspace{0.05in}

%\textbf{(2) Regulatory Mechanism Modeling and Learning.} 
To capture the cell response to environmental perturbation, a Michaelis–Menten (MM) formalism based regulation model was used to characterize the relationship of metabolic flux rates depending on the concentrations of associated substrates and inhibitors. This allows leveraging the existing biology knowledge and facilitating the learning of regulation mechanisms of iPSC metabolic reaction network from the experimental data.

Specifically, the $g$-th flux rate at the time $t$ is modeled as,  
\begin{equation}
    v_g[\pmb{u}(t)] = v_{max,g} \prod_{y \in \Omega^{g}_Y} \frac{u_y(t)}{u_y(t) + K_{m,y}} \prod_{z \in \Omega^g_Z} \frac{K_{i,z}}{u_z(t) + K_{i,z}},
    \label{eq:flux}
\end{equation}
for $g = 1, 2, \ldots, n$, where the set $\Omega^{g}_Y$ represents the collection of activators (such as nutrients and substrates) influencing the flux rate $v_g$ and the set $\Omega^g_Z$ represents the collection of inhibitors dampening $v_g$. The parameters $K_{i,\cdot}$, $K_{m,\cdot}$, and $v_{max,\cdot}$ represent the affinity constant, the inhibition constant, and the maximum specific flux rate respectively. For iPSC cultures, the experimental data referenced were used to identify the critical activators and inhibitors, as well as learning the MM regulation model coefficients. 

\vspace{0.05in}

\textbf{(2) Time-Course Isotopic Labeling Simulation.} 
Different $^{13}$C-labeling patterns are generated by different flux distributions $\pmb{v}[\pmb{u}(t)]$. Thus, incorporating the dynamic system of isotopic labeling patterns (MID), which is a vector containing the fractional abundance of each mass state of metabolites, into the metabolic network kinetic model can aid in understanding intercellular metabolic network mechanisms.
%The $^{13}$C-labeling patterns are highly dependent on relative pathway fluxes $\pmb{v}[\pmb{u}(t)]$, i.e. different labeling patterns will be produced for different flux distributions. Incorporating the dynamic model of the stable isotopic labeling system can facilitate the learning of intercellular metabolic network mechanisms.  
 %As such, fluxes can be inferred from measured isotopic labeling patterns.  The MID is a vector that contains the fractional abundance of each mass state of an EMU. 
%\cite{antoniewicz2007elementary} present a framework for modeling isotopic tracer systems that significantly reduces the number of system variables without any loss of information. 
%The elementary metabolite units (EMU) framework is a bottom-up modeling approach that is based on a highly efficient decomposition algorithm that identifies the minimum amount of information needed to simulate isotopic labeling within a reaction network. see more detailed information in studies, such as \cite{ahn2011metabolic,ahn2013parallel,antoniewicz2015methods,antoniewicz2021guide,antoniewicz2007elementary,antoniewicz2018guide,jazmin2013isotopically}. 
However, modeling each individual atom as one system state variable is computationally expensive. To address this issue, the elementary metabolite unit (EMU) framework was proposed, and it is based on a highly efficient decomposition method that can identify the minimum amount of information needed to simulate isotopic labeling \citep{antoniewicz2007elementary}. Basically, the EMUs are created by using a decomposition algorithm and form the new basis for generating system equations that describe the relationship between fluxes and isotope measurements; see more detailed information in studies\cite{ahn2011metabolic,ahn2013parallel,antoniewicz2015methods,antoniewicz2021guide,antoniewicz2007elementary,antoniewicz2018guide,jazmin2013isotopically}.

Based on the study \citep{antoniewicz2007elementary,young2008elementary}, the reduced system can be obtained after decoupling 
based on EMUs with size $r = 1, 2, \dots, R$ and connectivity. 
The time-dependent network was first identified: the decoupled EMUs with size $r$ network, $G_r(t) = \{V_r, E_r, W_r(t)\}$ with $r = 1, 2, \dots, R$. The vector $V_r = \big ({V_r^{(a)}}^\top, {V_r^{(b)}}^\top \big)^\top$ is the set of vertices (i.e., EMUs) within the $r$-th network, the vector $V^{(b)}_r$ with dimension $|V_r^{(b)}|$ is the set of input EMUs (i.e., EMUs with size smaller than $r$ or EMUs of extracellular carbon sources), the vector $V^{(a)}_r$ with dimension $|V_r^{(a)}|$ is the set of EMUs with size $r$. %Denote the number of EMUs of $V_r^{(a)}$ and $V_r^{(b)}$ as $n(V_r^{(a)})$ and $n(V_r^{(b)})$ respectively. 
$E_r$ is the adjacency matrix with dimension $\big(|V_r^{(a)}| + |V_r^{(b)}| \big) \times \big( |V_r^{(a)}| + |V_r^{(b)}| \big)$ representing the dependence between each vertex. The corresponding weight matrix $W_r(t)$ varies with time $t$. The non-negative $(i,j)$-th element $W_r^{(i,j)}(t)$ indicates the flux rate of the reaction producing $i$-th EMU by consuming $j$-th EMU at time $t$.

%(2) the input network $G^{in}_r(t) = \{V^{in}_r, E^{in}_r, W^{in}_r(t)\}$ to $G_r(t)$, where $V^{in}_r = \{V^{in(a)}_r, V^{in(b)}_r \}$ is the set of input EMUs $V^{in(a)}_r$ and their adjacent vertices $V^{in(b)}_r$ within $G_r(t)$, $E^{in}_r$ is the $n(V^{in(b)}_r) \times n(V^{in(a)}_r)$ adjacency matrix representing the dependence between the input EMUs and their adjacent vertices, and $W^{in}_r(t)$ is the flux rate of the corresponding reactions at time $t$.
%is the collection of input EMUs and/or carbon source EMUs of $r$-th decoupled reaction network and matrix $V_r$ represents $r$-th network connectivity. The non-negative element $V_r(i, j)$ indicates the flux rate of the reaction consuming $i$-th EMU and producing $j$-th EMU. 

Thus, at any time $t$, the dynamic isotopic labeling system can be defined as:
\begin{equation}
   { \frac{dC_r(t)}{dt}=\frac{A_r(t) \cdot C_r(t)+B_r(t) \cdot D_r(t)}{P_r(t)},}
\end{equation}
where the rows of the state matrix $C_r(t)$ correspond to the MIDs of EMUs within $V^{(a)}_r$ at time $t$. %{(MID are positive thus, the negative sign to represents the consume)} 
The input matrix $D_r(t)$ is analogous but with rows of input/carbon sources EMUs within $V^{(b)}_r$ at time $t$. 
%These MIDs are either calculated via $1$ to  $r-1$-th blocks or known as carbon sources. 
The concentration matrix $P_r(t)$ is a diagonal matrix whose elements are pool sizes corresponding to EMUs represented in $V_r^{(a)}$. The construction of $A_r(t)$ and $B_r(t)$ are based on the decoupled EMU reaction network $G_r(t)$. % and the input network $G^{in}_r(t)$. 
The system matrix $A_r(t)$ with size $|V^{(a)}_r| \times |V^{(a)}_r|$ and matrix $B_r(t)$ with size $|V^{(a)}_r| \times |V^{(b)}_r|$ describe the metabolic network with elements defined as follows: %{eq.(5) needs to be rewritten. Just online with rigorous math description}

\begin{equation}
\label{equ:Amatrix}
A_{r}^{(i,j)}(t) =\left\{
\begin{aligned}
- \sum_{k=1}^{|V_r^{(a)}|} W_r^{(k, j)}(t)  , &~~~ i = j, \\
W_r^{(i, j)}(t) , &~~~ i \neq j;
\end{aligned}
\right.
\end{equation}
and
\begin{equation}
    \label{equ: Bmatrix}
    \begin{aligned}
        B_r^{(i,j)}(t) = W_r^{(i, |V_r^{(a)}| + j)}(t)
    \end{aligned}.
\end{equation}
%\mbox{\vbox{\noindent time $t$ flux of $i$-th EMU in $X_r$\\from $j$-th EMU in $D_r(t)$}}

\subsection{iPSC Metabolic Reaction Network }
\label{subsec:reactionnetwork}

\begin{figure}[h!]
    \centering
    \includegraphics[width=0.5\textwidth]{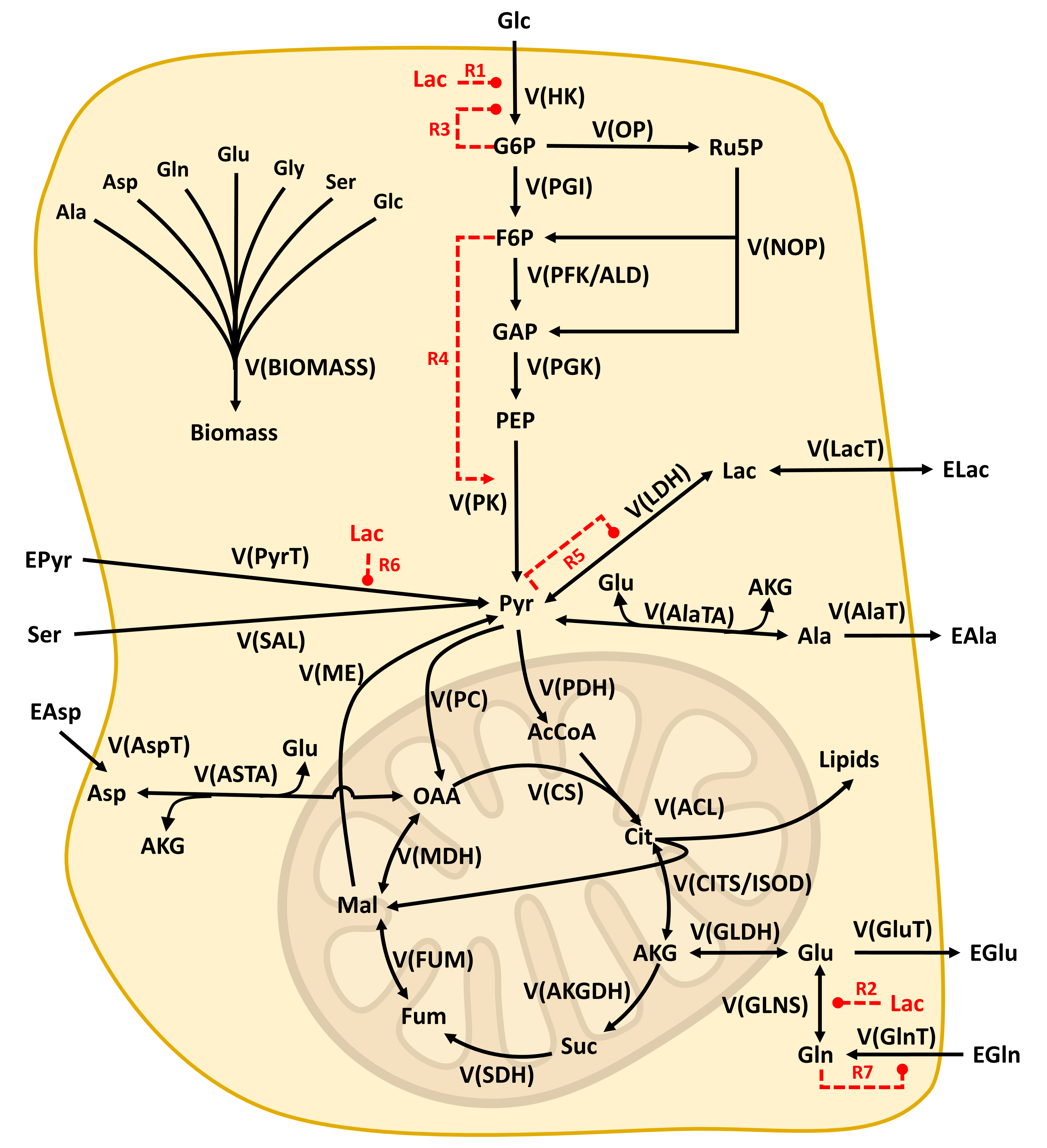}
    \caption{
    {
    An illustration of the iPSC regulatory metabolic network (Created with BioRender.com). Glycolysis, PPP, TCA and anaplerosis and amino acid utilization are shown. Additionally, the regulatory mechanisms included in the dynamic model are shown. The metabolites descriptions are listed in Table~\ref{tab:metabolite}. The enzyme description, including the  Enzyme Commission Numbers (EC-No.) for each reaction,  is listed in Table~\ref{tab:enzyme}.} 
    }
    \label{fig:network}
\end{figure}

\begin{sloppypar}
The developed iPSC metabolic network model %incorporated 
leverages central carbon metabolism from several previously published metabolic frameworks \citep{dressel2011effects,fiorenza2020value,ghorbaniaghdam2014silico,niklas2012metabolic,ghorbaniaghdam2013kinetic,odenwelder2021induced} and further learns from data. The iPSC metabolic network included glycolysis, TCA cycle, anaplerosis, PPP, and amino acid metabolism. For simplification, the reactions of PPP were collapsed into two reactions: Oxidative phase/branch and Non-oxidative phase/branch (i.e., No. 9 \& 10 reaction in Table~\ref{tab:reaction}). During the exponential phase, greater than 90\% of pentose-phosphate carbons is returned to glycolysis for the mammalian cell \citep{templeton2013peak,halestrap2012monocarboxylate}. Thus, the synthesis of nucleotides and nucleic acids is considered as an insignificant contributor to the model. 
The metabolic network is shown in Fig.~\ref{fig:network} and the metabolic stoichiometry is listed in Table~\ref{tab:reaction}. The descriptions of metabolites and the enzymes conform to the Enzyme Commission Number (EC-No.) and are provided in Tables~\ref{tab:metabolite} and ~\ref{tab:enzyme}, respectively.

{While this model primarily focuses on central carbon metabolism, it's worth noting that the additional model's structure, including metabolic reaction pathways (e.g., one-carbon metabolism, fatty acid oxidation, and nucleotide biosynthesis) and the associated flux regulatory mechanisms, could be easily incorporated.} 
\end{sloppypar}

\subsection{Biokinetic Model of Flux Regulatory Mechanism}
\label{subsec:regulatorymechanism}

\begin{sloppy}
Due to the lack of energetic state and redox level measurements, each metabolic flux rate is modeled as dependent on the substrates and inhibitors concentrations using Michaelis–Menten model formalism. Similar to the existing studies \citep{dressel2011effects,fiorenza2020value,ghorbaniaghdam2014silico}, for simplification purposes and also due to a lack of available data in the literature, a single affinity constant value is used for each metabolite.  Additionally, reaction reversibility was considered during the iPSC culture model development for some key reactions. The final model is fully described in Table~\ref{tab:kinetic}. Several regulatory mechanisms (i.e., No. \textbf{R1} to \textbf{R7}) are highlighted in Fig.~\ref{fig:network}, which were incorporated into the metabolic flux kinetics model to improve the model’s prediction capability and better characterize the responses to environmental variations.  These key final reactions are described below, where these new dependencies are highlighted in under brackets $\underbrace{\dots\dots}$ with the corresponding regulatory mechanism No., while the original nomenclature is outside the brackets.  %{modify}%{make the locations of R: consistent}
\end{sloppy}

\begin{itemize}
    \item Lactate accumulation has previously been reported to reduce glycolytic activity by inhibiting hexokinase (HK) and phosphofructokinase (PFK) activity in mammalian cells, where lactate acts as a signaling molecule to down-regulate PFK activity \citep{ivarsson2015insights,mulukutla2012metabolic,costa2007lactate}. After evaluation of the experimental data, the model for HK was updated to include this inhibitory effect of lactate on it: 
    %\begin{fleqn}
    \begin{equation}   \label{equ:R1} %\tag{\mbox{\textbf{R1}}}
    \begin{split}    
    %\mbox{\textbf{R1}:~~} 
    v(HK) = & v_{max, HK} \times \frac{Glc}{K_{m, Glc}+Glc} \\ & \times \underbrace{\frac{K_{i,LactoHK}}{K_{i, LactoHK}+Lac}}_{\textbf{R1}}.
    \end{split}
    \end{equation}
    %\end{fleqn}
    
    \item Since lactate inhibits glutaminase activity -- the enzyme responsible for converting glutamine (GLN) to glutamate (GLU) \citep{glacken1988mathematical,hassell1991growth} -- the forward ($f$) flux rate for the reaction, i.e., Gln $\leftrightarrow$ Glu + NH$_4$, was updated, 
    %\begin{fleqn}
    \begin{equation} \label{equ:R2} 
    \begin{split}
     %\mbox{\textbf{R2}:~~}    
     v(GLNSf) = & v_{max, fGLNS} \times \frac{Gln}{K{m, Gln}+Gln} \\ & \times 
          \underbrace{\frac{K_{i, LactoGLNS}}{K_{i, LactoGLNS}+Lac}}_{\textbf{R2}}.
    \end{split}     
    \end{equation}
    %\end{fleqn}
    \item Several regulatory functions, adapted from the studies\cite{ghorbaniaghdam2013kinetic,ghorbaniaghdam2014analyzing,ghorbaniaghdam2014silico}, were evaluated using the experimental data to characterize activations and inhibitions. The regulatory mechanisms involved in glycolysis are described as: a) hexokinase inhibition by its product G6P, see \textbf{R3} in eq~(\ref{equ:R3}); 
    b) activation of pyruvate kinase by F6P, see \textbf{R4} in eq~(\ref{equ:R4}), as well as c) the inhibition of lactate dehydrogenase reverse ($r$) reaction, see \textbf{R5} in eq~(\ref{equ:R5}), was changed to include the G6P inhibition: 
    %\begin{fleqn}
    \begin{equation} \label{equ:R3} 
    \begin{split}
    %\mbox{\textbf{R3}:~~}    
    v(HK) = & v_{max, HK} \times \frac{Glc}{K_{m, Glc} +Glc }  \\ & \times \frac{K_{i, LactoHK}}{K_{i, LactoHK}+Lac} \\ & 
    \times \underbrace{\frac{K_{i, G6P}}{K_{i, G6P}+G6P}}_{\textbf{R3}}; 
    \end{split}     
    \end{equation}    
    %\end{fleqn}
    
    %\begin{fleqn}
    \begin{equation} \label{equ:R4} 
    \begin{split}
     %\mbox{\textbf{R4}:~~}   
     v(PK) = & v_{max, PK} \\& \times \frac{PEP}{K_{m, PEP}  \times \underbrace{\Big[1+\frac{K_{a, F6P}}{F6P} \Big]}_{\textbf{R4}}+PEP}; 
    \end{split}     
    \end{equation}
    %\end{fleqn}
    
    %\begin{fleqn}
    \begin{equation} \label{equ:R5} 
    \begin{split}
    %\mbox{\textbf{R5}:~~}    
    v(LDHr)= & v_{max, rLDH} \times \frac{Lac}{K_{m, Lac}+Lac} \\ &  \times  \underbrace{\frac{K_{i, Pyr}}{K_{i, Pyr}+Pyr}}_{\textbf{R5}}. 
    \end{split}     
    \end{equation}
    %\end{fleqn}
    
    \item Both lactate and pyruvate transport across the plasma membrane are facilitated by proton-linked monocarboxylate transporters (MCTs) \citep{halestrap2012monocarboxylate,huckabee1956control}. More favorable lactate transport kinetics may decrease pyruvate consumption under high lactate culture conditions \citep{odenwelder2021induced,draoui2011lactate}. Thus, the model of PyrT was updated to:
    %\begin{fleqn}
    \begin{equation}  \label{equ:R6} 
    \begin{split}
    %\mbox{\textbf{R6}:~~}    
    v(PyrT) = & v_{max, PyrT} \times \frac{EPyr}{K_{m, EPyr}+EPyr} \\ & \times \underbrace{\frac{K_{i, LactoPyr}}{K_{i, LactoPyr}+Lac}}_{\textbf{R6}}. 
    \end{split}     
    \end{equation}
    %\end{fleqn}
    \item Since the transportation of extracellular-glutamine via cell membrane can be inhibited by intracellular-glutamine, the model for GlnT was updated to:   
    %\begin{fleqn}
    \begin{equation} \label{equ:R7} 
    \begin{split}
    %\mbox{\textbf{R7}:~~}    
    v(GlnT) = & v_{max, GlnT} \times \frac{EGln}{K_{m, EGln}+EGln} \\ & \times \underbrace{\frac{K_{i, GLN}}{K_{i, GLN}+GLN}}_{\textbf{R7}}. 
    \end{split}     
    \end{equation}
    %\end{fleqn}
\end{itemize}

% >>>>

\subsection{Model Fit and Goodness-of-fit Assessment}
\label{subsec: goodness-of-fit}

%For biochemical processes, the laws of biophysical chemistry often allow us to construct mechanistic models. While these models can describe the the mean changes in a reaction profile over time, 

\begin{sloppypar}

The proposed iPSC metabolic network kinetic model focuses on characterizing central carbon metabolism, including significant reactions from glycolysis, the TCA cycle, anaplerosis, PPP, and amino acid metabolism, along with regulatory mechanisms, across distinct cultural environments characterized by varying levels of glucose and lactate concentrations. %It is well known that in iPSC aggregate cultures, the glucose concentration will be low at the aggregate center and the lactate concentration will be high \citep{kinney2011multiparametric}, as its proper functioning is critical for the survival and proliferation of iPSCs \citep{nelson2008lehninger}, as well as cell product quality assurance.
%In iPSC aggregate cultures, the well-established pattern involves low glucose concentration at the aggregate center and high lactate concentration \citep{kinney2011multiparametric}. Understanding the implications of these concentration variations is crucial for the survival and proliferation of iPSCs \citep{nelson2008lehninger}, and also essential for maintaining the quality assurance of cell products.} 
The final model developed is shown in Table~\ref{tab:reaction}.  This model includes 30 metabolic reactions with 32 variables (metabolites' concentration) in the reaction equations. {The iPSC metabolic flux rate regulation biokinetic model, presented in Table~\ref{tab:kinetic}, incorporates key activators and inhibitors for each reaction, describing the regulatory effects through the Michaelis–Menten model.}%The kinetic formulations for the flux regulation are presented in Table~\ref{tab:kinetic}. %which includes 57 kinetic parameters: 20 saturation constants (K’s), 30 maximum reaction constants (vmax’s), and one parameter for each regulatory function (unclear). 

The proposed iPSC culture kinetic model estimates model parameters using {both} extracellular and MID data collected over time under the four different culture conditions. %Denote the available extracellular metabolites concentrations data at time $t$ as $s_t^{meas}$ for $t=0,12,24,36,48$ hour. 
{The extracellular metabolite concentration measurements at time $t$ are denoted as $s_t^{meas}$ for $t=0, 12, 24, 36, 48$ hours (h). To ensure the model's accuracy and minimize the impact of measurement errors, values that fell below the detection limit of the Cedex Bioanalyzer are treated as missing data. The intracellular isotopic labeling measurements at time $t^\prime$ are denoted as $\mbox{MID}_{t^\prime}^{meas}$ for $t^\prime=24$ and $48$ hours. The MID measurements obtained from [U-$^{13}$C$_3$] lactate for the high glucose high lactate and low glucose high lactate cultures exhibited considerable measurement errors, as reported in the study \cite{odenwelder2021induced}. Consequently, these measurements were intentionally excluded from the training dataset used for developing the proposed model. } %Denote intracellular isotopic labeling measurements at time $t^\prime$ as $\mbox{MID}_{t^\prime}^{meas}$ for $t^\prime=24,48$ hour. 

{
For model fitting, the objective is to minimize the weighted sum of squared residuals (SSR) between the available experimental data, which includes both extracellular metabolite concentrations ($s_t^{meas}$) and intracellular isotopic labeling ($\mbox{MID}_{t^\prime}^{meas}$), and the corresponding model-predicted values ($s_t^{sim}$ and $\mbox{MID}_{t^\prime}^{sim}$),
}%For the model fitting, the objective is minimizing the weighted sum of squared residuals (SSR) between available experimental data ($s_t^{meas}$, $\mbox{MID}_{t^\prime}^{meas}$) and model predicted values ($s_t^{sim}$, $\mbox{MID}_{t^\prime}^{sim}$), }
%\vspace{-0.38in}
\begin{equation}
    \min \mbox{SSR} = \sum_{t=1}^T \frac{(s_t^{sim} - s_{t}^{meas})^2}{var_t^s} + \sum_{t^\prime=1}^{T^\prime}\frac{(\mbox{MID}_{t^\prime}^{sim} - \mbox{MID}_{t^\prime}^{meas})^2}{var_{t^\prime}^M},
    \label{sq.SSR}
\end{equation}
{where the weights are the inverse of the variances of the experimental measurements on extracellular metabolites and MIDs, i.e., ($var_t^s$, $var_{t^\prime}^M$). The fitting criteria in eq~(\ref{sq.SSR}) allows the model to systematically and effectively fit the metabolic regulation network at different times.}
%the weights are the inverse of the variances of the experimental measurement ($var_t^s, var_{t^\prime}^M$) so that the model can systematically fit the metabolic regulation network well at different times. 

The initial conditions for extracellular metabolites were obtained from culture data. %As the experimental data only include the relative abundance and the MIDs, estimates for the initial intracellular metabolite concentrations were sourced from the literature for mammalian cells cultured under similar conditions (\cite{chang2021brenda} and references therein). The initial kinetic parameter values were taken from literature for similar metabolic networks and pathways (\cite{chang2021brenda} and references therein). 
As the experimental data only includes the relative abundance and the MIDs, estimates for the initial intracellular metabolite concentrations were sourced from the BRENDA Enzyme Database \cite{chang2021brenda} and related references for mammalian cells cultured under similar conditions. Similarly, the initial kinetic parameter values were taken from the BRENDA Enzyme Database \cite{chang2021brenda} and relevant references for similar metabolic networks and pathways.

The goodness-of-fit for the developed iPSC metabolic kinetic model was assessed based on predictions using a chi-square test. Basically, the null hypothesis is that the fitted model can faithfully represent the iPSC culture metabolic mechanism. This test assumes that the minimized variance-weighted SSR follows a chi-square distribution with $d$ degrees of freedom, where the degree of freedom $d$ equals the number of observations minus the number of fitted parameters.  In this study, {the chi-square statistical test has $p$-value much greater than $0.05$ indicating} the model predictions are not significantly different than the measured values.
%for the chi-square statistical test.
\end{sloppypar}

% =========================================

\section{Results and Discussion}
\label{sec:results}

{The proposed mechanistic model was %estimated with limited 
developed based on a small set of experimental data (see Section~\ref{sec:datadescription})
and the model predictive performance was then validated in two manners. 
First, to mimic the dynamic data collection and assess the rolling forecasts required for 
process control, at any time $t$,
historical data were used to fit the model up to $t$ and then the fitted model was used to predict the remainder of the culture behavior. }
In Section~\ref{subsec:iPSCPred}, the model's ability to capture the dynamic evolution of iPSC metabolic characteristics was evaluated. The model was trained on experimental data collected over different time intervals (0-h to 12-h, 0-h to 24-h, and 0-h to 36-h), and used to predict the iPSC culture for up to 48-h. 
%This approach mimics the dynamic data collection process and predict to the end of culture process in order to support strategic .

Second, to assess the model's ability to generalize and extrapolate to new {culture} conditions, three datasets, randomly selected from 
%HGLL, HGHL, LGLL, and LGHL,
(1)  high glucose and low lactate (HGLL); (2) high glucose and high lactate (HGHL); (3) low glucose and low lactate (LGLL); and (4) low glucose and high lactate (HGHL), 
were used to predict the fourth dataset with different initial conditions in Section~\ref{subsec:iPSCPredAcross}. Basically, {three datasets were used to train the model, and then the model was used to predict 
the fourth dataset behaviors.
%the behavior of the fourth dataset given the initial condition. %fourth leave-out data with different initial conditions was used as the test dataset to assess the model extrapolation prediction performance. 
This training/test dataset selection and evaluation strategy is built on the philosophy of cross validation (CV) that is often used in the literature on model selection  \cite{mclachlan2005discriminant,refaeilzadeh2009cross}.
Each  culture condition
%: (1)  high glucose and low lactate (HGLL); (2) high glucose and high lactate (HGHL); (3) low glucose and low lactate (LGLL); and (4) low glucose and high lactate (HGHL), 
was examined in succession using the other three datasets.} 
Therefore, by validating the model's prediction performance under different experimental conditions, % and initial states, 
its robustness and usefulness for predicting iPSC characteristics in a variety of settings is verified.

\subsection{Prediction of iPSC Culture Process}
\label{subsec:iPSCPred}
\begin{figure*}[h!]
    \centering
    \includegraphics[width=1\textwidth]{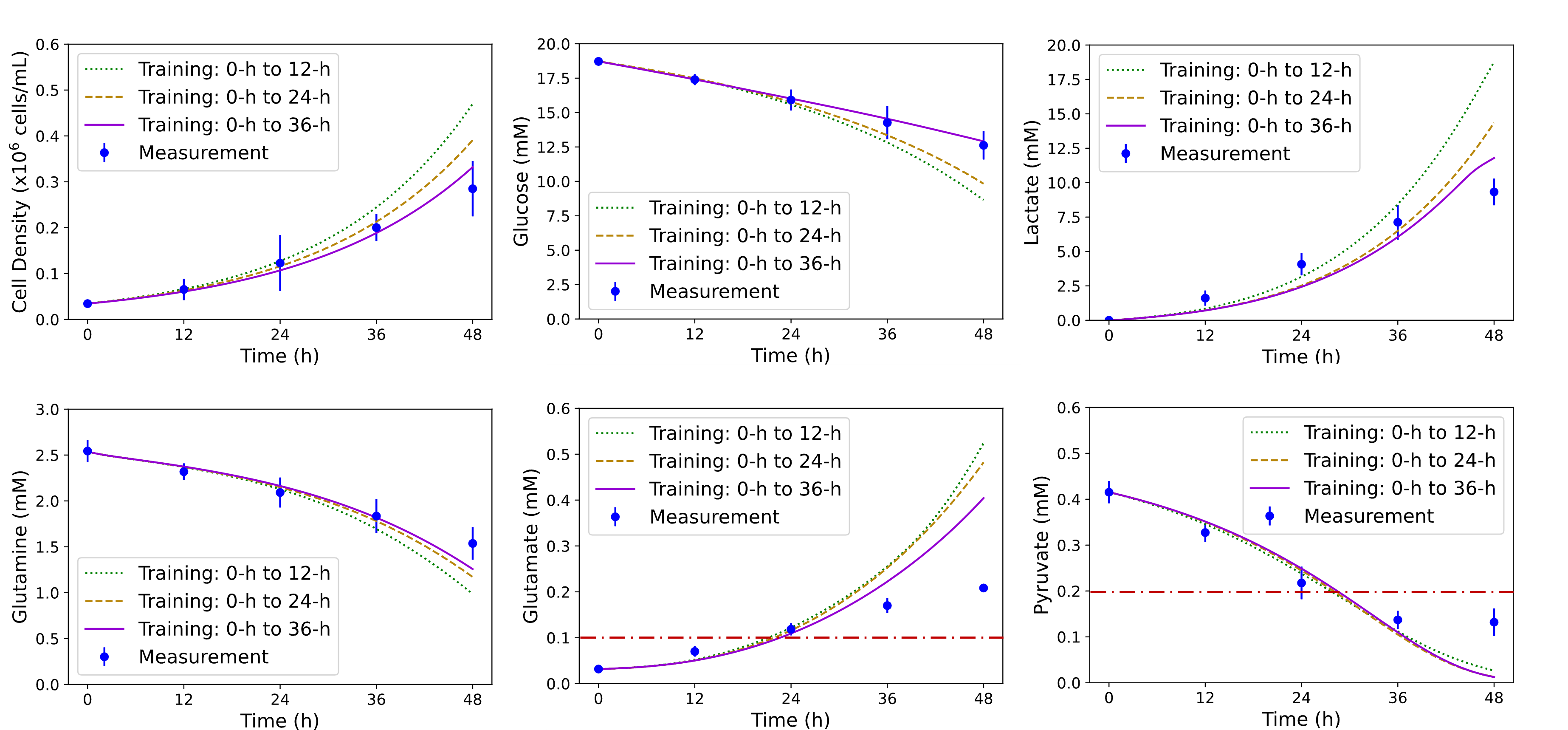}
    \caption{{Dynamic model trained with historical data collected under different time intervals - Cell characteristic predictions for high glucose and low lactate cultures. (A) Cell density, (B) Glucose, (C) Lactate, (D) Glutamine, (E) Glutamate, and (F) Pyruvate. Times 0-h to 12-h (green dotted line); Times 0-h to 24-h (brown dashed line); and Times 0-h to 36-h (purple solid line). The detection limit of the Cedex Bioanalyzer is shown as the red dash-dot line.  }} 
    \label{fig:pred_forward_hgll}
\end{figure*}

During the model validation process, the proposed model simulator was trained using experimental data collected over various time intervals (0-h to 12-h, 0-h to 24-h, and 0-h to 36-h). The model's ability to predict the dynamic trajectories of key metabolites, including glucose, lactate, glutamate, glutamine, and pyruvate, as well as cell density, was evaluated by comparing the predicted values to experimental observations up to 48 hours.
Fig.~\ref{fig:pred_forward_hgll} provides a representative result of the cell culture process
prediction for the case starting with HGLL. The results for the remaining settings, including HGHL, LGLL, and LGHL, are presented in Appendix Fig.~\ref{fig:pred_forward_hghl} to \ref{fig:pred_forward_lghl}. 

{The Chi-square test was employed to assess the goodness-of-fit of the proposed mechanistic model trained using the historical data collected in different time intervals. The statistical test for the model trained on 0-h to 12-h of experimental data yielded a value of 33.6 with 47 degrees of freedom ($p$-value = 0.93). Similarly, for the model trained on 0-h to 23-h and 0-h to 36-h of experimental data, the test statistics was 82.6 with 107 degrees of freedom ($p$-value = 0.96) and 138.1 with 167 degrees of freedom ($p$-value = 0.95), respectively. 
All of the p-values exceed 0.05, suggesting that the proposed metabolic kinetics model fits well with the experimental dataset. The results in Fig.~\ref{fig:pred_forward_hgll} demonstrate that the fitted model can provide improved predictions for most of the metabolite trajectories as more data are collected over time.}
%The chi-square tests conducted on different time intervals were used to assess the goodness-of-fit of the predictive models. The prediction test statistics for the model trained on 0-12 hours of experimental data was 33.6 with 47 degrees of freedom ($p$-value = 0.93), while the test statistics for the model trained on 0-24 hours and 0-36 hours of experimental data were 82.6 with 107 degrees of freedom ($p$-value = 0.96) and 138.1 with 167 degrees of freedom ($p$-value = 0.95), respectively. 
%All of the p-values are much greater than 0.05, which indicates the proposed metabolic kinetics simulator fits well with the experimental dataset. The results in Fig.~\ref{fig:pred_forward_hgll} demonstrate that the fitted model can provide better predictions for most of the metabolite trajectories as more data are collected in time. 

Overall, the predicted profiles of the iPSC cultures can closely track the dynamic patterns of the measured profiles and capture the cell culture dynamics. The experimental iPSC data were obtained from plates and dishes with limited space. Even though the overall growth rates were the same across the four conditions by experimental design, the growth rate decreased with time as nutrients were consumed and metabolic wastes accumulated. It is important to note that the developed mechanistic metabolic kinetic model is a simplified representation of the iPSC carbon central network, there could likely be some key metabolic regulations and biochemical reactions that were not incorporated. %; but they might be included for aggregate cultures. 
The less accurate predictive 
%lower  predictive 
performances for lactate (under low lactate conditions) and glutamate may be due to the omission of key reactions or regulatory functions. For example, as suggested by the calibrated model in previous studies \citep{ghorbaniaghdam2013kinetic,ghorbaniaghdam2014analyzing,ghorbaniaghdam2014silico}, the membrane transportation of glutamate for mammalian cells is significantly impacted by the energetic state (ATP and ADP), and the forward and reverse conversion of lactate to pyruvate is influenced by the redox level (NAD and NADH). 

Since the redox level and energetic state-related measurements were not collected for the iPSC cultures, these factors were not accounted for in the current model. 
The study\cite{nolan2011dynamic} found for CHO cells, a redox parameter assisted with lactate predictions; however, the redox parameter implemented by this study was unmeasurable and thus would be difficult to incorporate into a control model. Furthermore, the lower prediction performance for glutamate may be attributed to its involvement in multiple reactions (see Fig.~\ref{fig:network}), leading to error accumulation over time. Future work should address these limitations with more comprehensive experimental data for iPSC cultures.

% ==================================

\subsection{Prediction across Different Initial Conditions}
\label{subsec:iPSCPredAcross}

{%To assess the predictive extrapolation capabilities of the proposed simulator across various initial culture conditions and enhance understanding of iPSCs' responses to changes in culture conditions, particularly variations in glucose and lactate levels, a cross-validation approach was employed. This method entailed training the model with data from three out of four datasets while reserving one dataset for evaluation. 
To assess the %extrapolation 
predictive capability of the proposed mechanistic model across various initial culture conditions and enhance the understanding of iPSCs response to environmental changes, specifically glucose and lactate, a cross-validation approach was employed. % to assess the model prediction performance. In specific, 
This approach first trained the model with three of the four datasets 
(i.e., HGLL, HGHL, LGLL, and HGHL).
The remaining dataset behavior was then predicted. %used for the evaluation of model prediction performance. 
This approach provided valuable insight for optimizing Design of Experiments (DoE) strategies in the realm of \textit{in silico} experimentation.}

{
The model %extrapolation predictions for 
cross-validation prediction for the HGLL culture was evaluated using the model trained on HGHL, LGLL, and LGHL datasets, as illustrated in Fig.~\ref{fig:pred_across_hgll}. The other three across-validation predictions are provided in Appendix Fig.~\ref{fig:pred_across_hghl} to \ref{fig:pred_across_lghl}. }
%In order to understand how iPSC responds to different cultural conditions (i.e., variable glucose and lactate), cross validation was used, where one dataset is left out while training the model with data from the remaining three datasets.  Fig.~\ref{fig:pred_across_hgll} shows the results of this {cross validation} approach for the HGLL cultures where the  HGHL, LGLL, and LGHL datasets were used. 
Cell density and key metabolites (i.e., glucose, lactate, glutamate, glutamine, and pyruvate) were predicted and are shown with the average of the measured values with standard deviations. %The across-condition extrapolation predictions for other culture conditions are provided in Appendix Fig.~\ref{fig:pred_across_hghl} to \ref{fig:pred_across_lghl}. 
To evaluate the goodness-of-fit, the chi-square test SSR statistics is 783 with 802 degrees of freedom with $p$-value = 0.67 much greater than 0.05. This p-value indicates that the metabolic kinetic model can faithfully represent the iPSC culture regulatory mechanisms under different levels of glucose and lactate concentrations. 

\begin{figure*}[h!]
    \centering
    \includegraphics[width=1\textwidth]{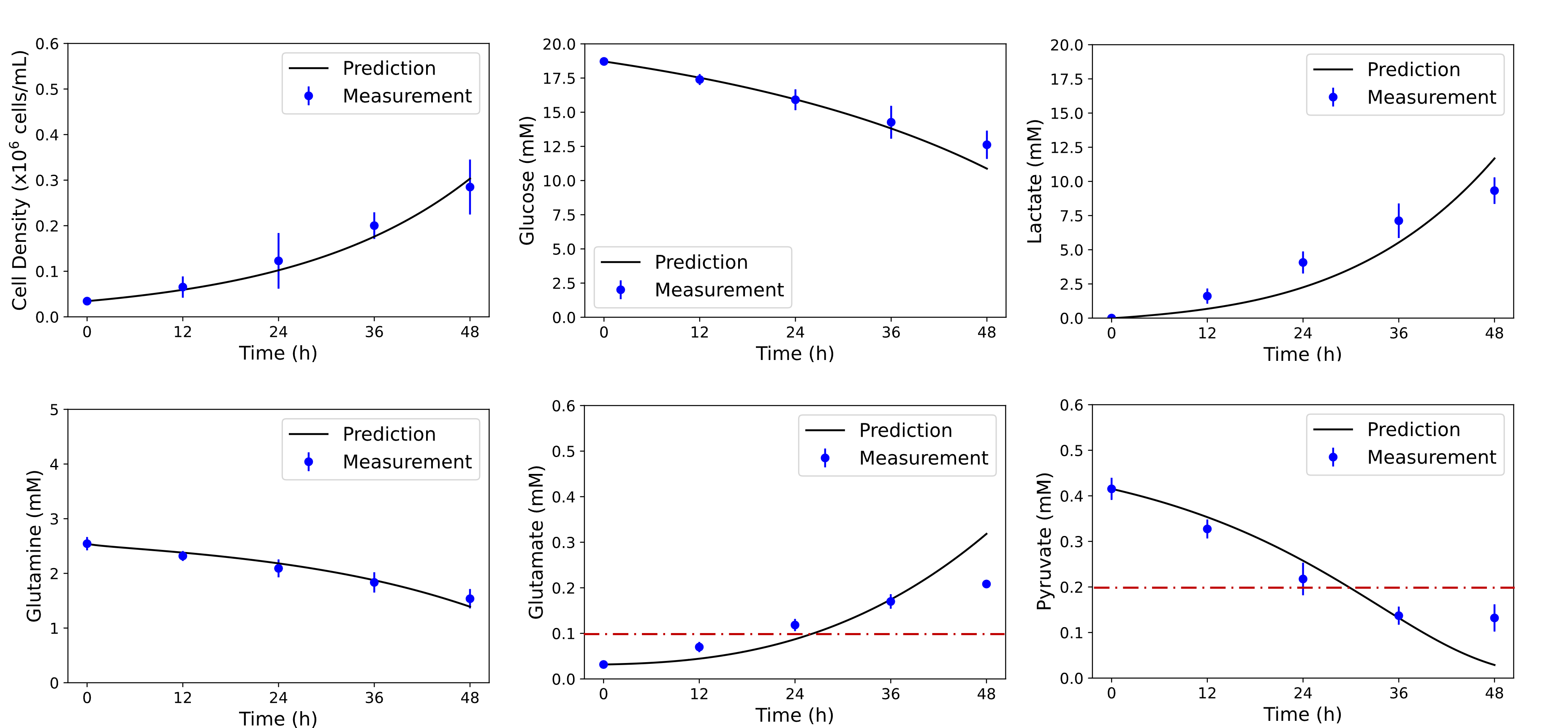}
    \caption{{Dynamic model trained with three other datasets with varied initial glucose and lactate concentrations for cell characteristic predictions in high glucose and low lactate cultures.  (A) Cell density, (B) Glucose, (C) Lactate, (D) Glutamine, (E) Glutamate, and (F) Pyruvate. The detection limit of the Cedex Bioanalyzer is shown as the red dash-dot line.}} \label{fig:pred_across_hgll}
\end{figure*}

\begin{figure}[h!]
    \centering
    \includegraphics[width=0.5\textwidth]{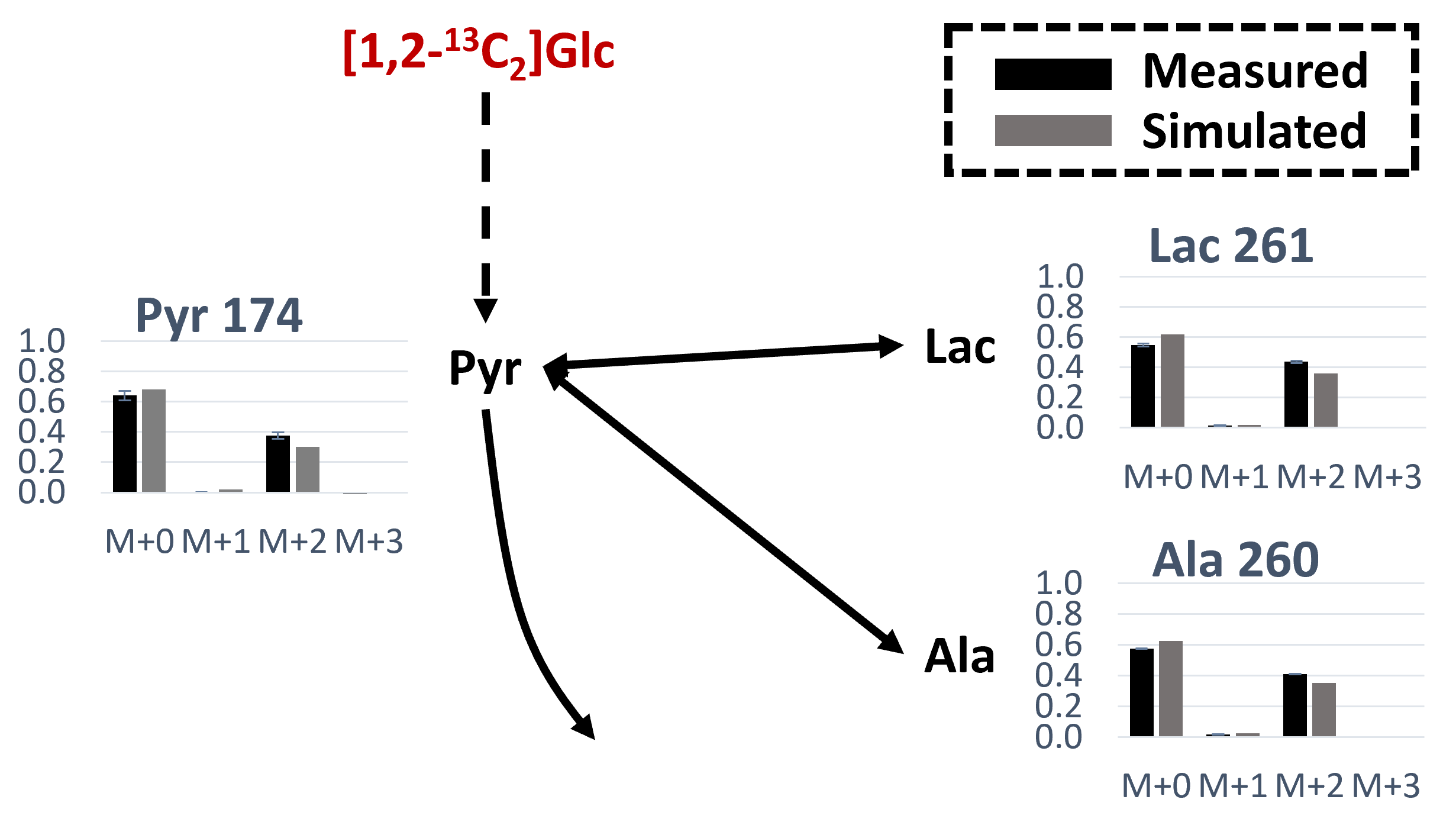}
    \caption{{MID measurements and predictions for the [1,2-$^{13}$C$_2$] glucose tracer at 48-h for high glucose low lactate cultures using the dynamic model trained on the other three data sets compared to literature measurements.  MIDs shown have been corrected for natural abundance.}} 
    \label{fig:glc_control_48}
\end{figure}

\begin{figure}[h!]
    \centering
    \includegraphics[width=0.5\textwidth]{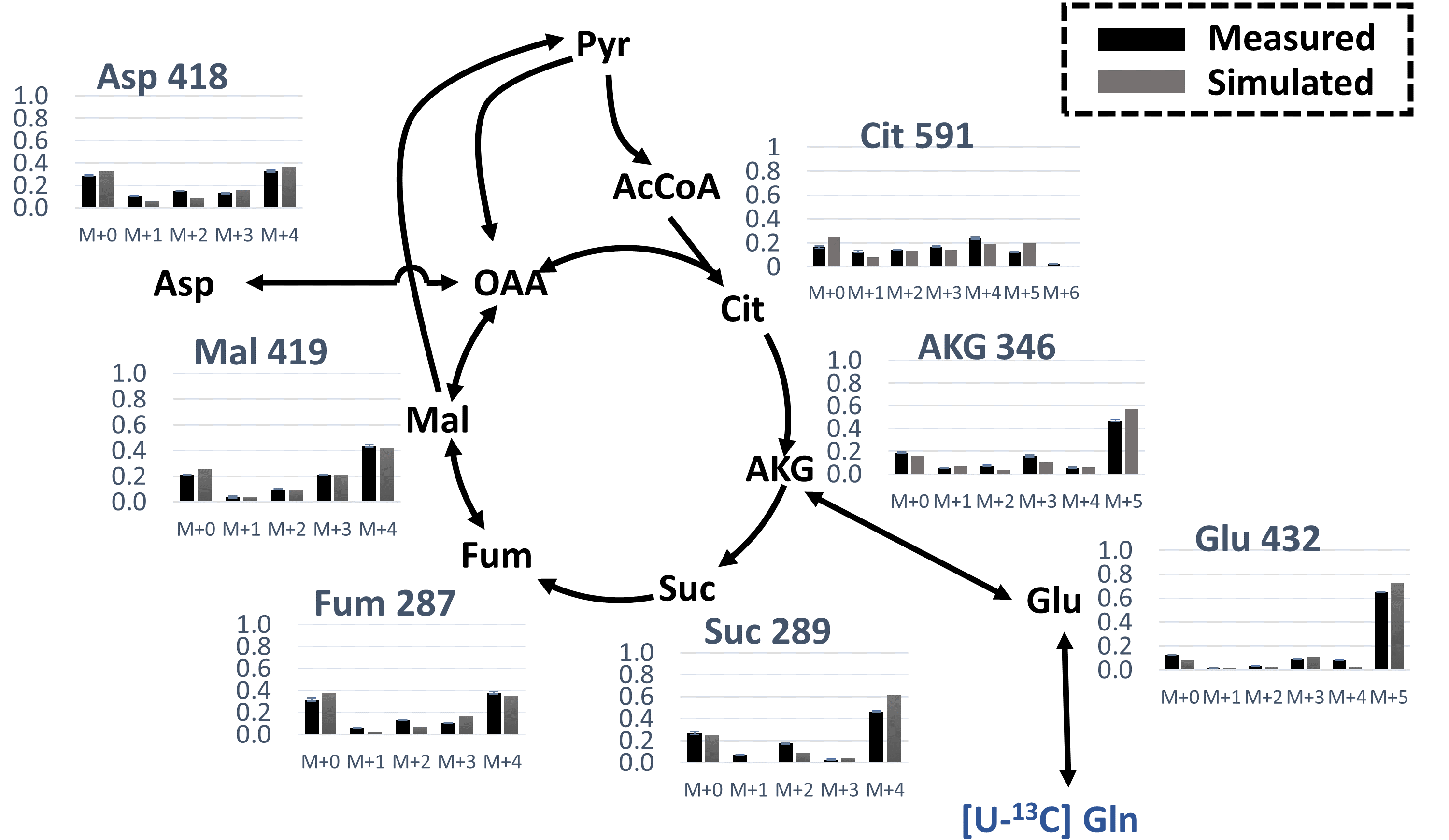}
    \caption{{MID measurements and predictions for the [U-$^{13}$C$_5$] glutamine tracer at 48-h for high glucose low lactate cultures using the dynamic model trained on the other three data sets compared to literature measurements.  MIDs shown have been corrected for natural abundance.}}
    \label{fig:gln_control_48}
\end{figure}

The predictions of intracellular MID from the [1,2-$^{13}$C$_2$] glucose and [U-$^{13}$C$_5$] glutamine tracer at 48-h under the control (HGLL) culture condition are shown in Fig.~\ref{fig:glc_control_48} and Fig.~\ref{fig:gln_control_48}, respectively. Even though there is some prediction error, the simulation model can correctly predict the dynamics and interdependencies of multivariate iPSC culture process 
metabolism. The developed metabolic kinetic model can faithfully predict the cell response to environmental perturbation, which could guide the strategic feeding strategy for the integrated iPSC culture process 
for future research. In addition, by incorporating even the simplified PPP reactions, the M$+$1 isotopic labeling pattern for pyruvate and related metabolites from [1,2-$^{13}$C$_2$] glucose tracer can be well-predicted. 

\begin{figure*}[h!]%
    \centering
    \subfloat{{\includegraphics[width=0.47\textwidth]{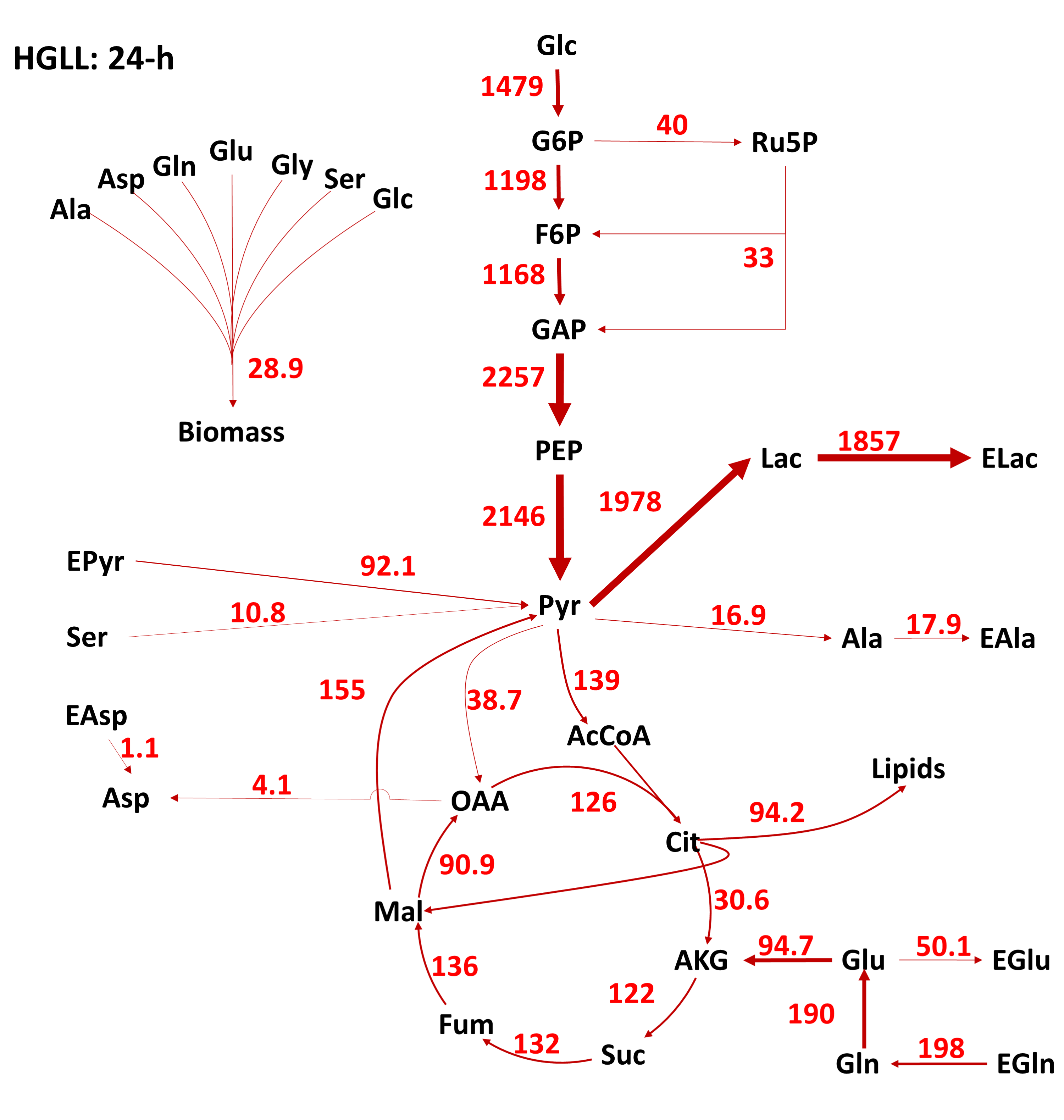} }}%
    \qquad
    \subfloat{{\includegraphics[width=0.47\textwidth]{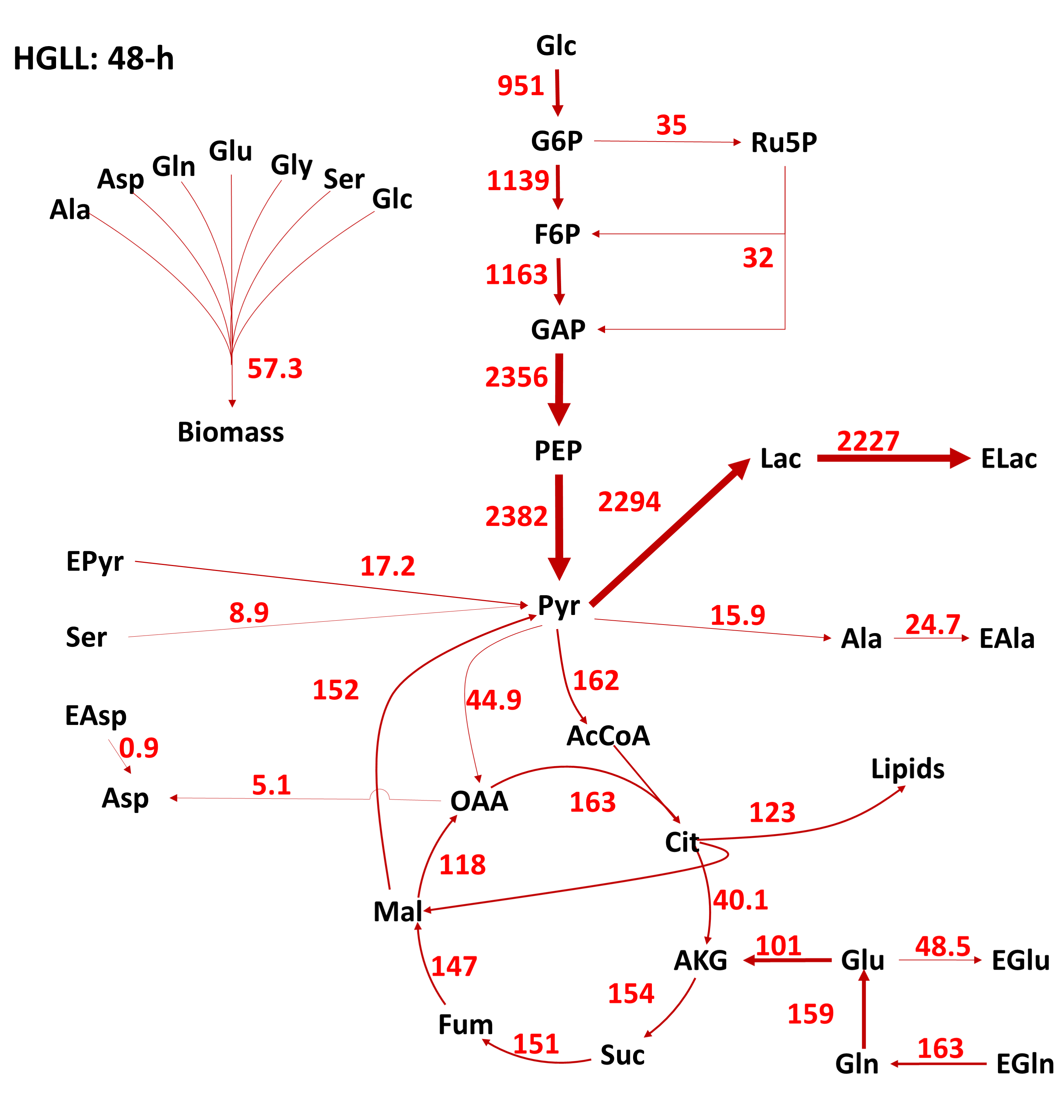} }}%
    \caption{Metabolic flux maps for K3 iPSC for the high glucose and low lactate cultures at 24-h and 48-h.  Predicted fluxes are given in nmol/10$^6$ cells$\cdot$h. The line thicknesses represent the relative fluxes.}%
    \label{fig:fluxmap_HGLL}%
\end{figure*}

The prediction of metabolic flux maps for K3 iPSC under control culture at 24-h and 48-h are shown in Fig~\ref{fig:fluxmap_HGLL}. The prediction performance of metabolic concentration trajectory and flux maps for K3 iPSC under HGHL, LGLL, and LGHL are provided in Appendix Fig.~\ref{fig:fluxmap_HGHL} to Fig.~\ref{fig:fluxmap_LGHL}. Since no restrictive steady-state assumption was required for the developed metabolic kinetic model, the flow-in flux is not required to be equal to the flow-out flux for each metabolite. 

{The prediction of metabolic flux dynamics for K3 iPSC culture reveals several significant observations. First, the glycolytic efficiency, which measures how effectively K3 iPSCs utilize glucose as an energy source, remained consistently high across all conditions, exceeding 1.6 and approaching the theoretical maximum of 2.0 moles of lactate produced per mole of glucose consumed.  This suggests that K3 iPSCs are efficiently converting glucose into lactate with minimal byproduct production. %wastage as other byproducts, indicating an effective utilization of glucose.
Second, the [U-$^{13}$C$_3$] lactate tracer showed that K3 iPSCs consumed and metabolized lactate in the high lactate cultures. However, in all four culture conditions, there was a net production of lactate, indicating that K3 iPSCs produced and also consumed lactate in the high lactate culture conditions. 
Third, in the control experimental setting, as the iPSC cultures approached the late exponential phase, there was a gradual increase in the flux rates of the TCA cycle, consistent with findings reported in a previous literature study \cite{templeton2013peak}. 
Finally, the mechanistic model successfully predicted the reduced glutamine consumption fluxes under high lactate culture conditions, along with a decrease in the conversion of glutamine to AKG. Consequently, the model also predicted slightly lower fluxes through key TCA cycle reactions such as $\alpha$-ketoglutarate dehydrogenase (AKGDH), succinate dehydrogenase (SDH), malate dehydrogenase (MDH), fumarase (FUM), and citrate synthase (CS). 
%These predictions align with the observations reported in the study \cite{odenwelder2021induced}. 
} 

{Table~\ref{tab:pred_flux} lists the predictions of biomass-specific uptake and production rates for critical extracellular metabolites at 12-h and 36-h. The table also includes the average flux between 12-h and 36-h, as determined by the Extracellular Time-Course Analysis (ETA) software \citep{murphy2013eta}, for comparison. The general mass balance equation, constructed by ETA, which describes the dynamics in the concentration of the $i$-th extracellular metabolite under batch growth conditions, is given by $$\frac{d{s_i}(t)}{dt} = r_i X(t),$$ where $s_i(t)$ represents the concentration of the $i$-th metabolite at time $t$, $r_i$ is the specific rate of metabolite consumption or production, and $X(t)$ stands for cell density at time $t$. This equation assumes no degradation of metabolites.
Notice that ETA results are based on the assumption of metabolic steady-state, meaning that the consumption or production rate $r_i$ of any $i$-th metabolite remains constant and it is not influenced by the changes in extracellular or intracellular concentrations. This assumption explains why most flux estimations using the ETA approach for 12-h to 36-h fall within the range of the model predictions at 12-h and 36-h. %using the developed metabolic kinetic model.
}
%\textcolor{blue}{(add a sentence to emphasize it also constraints the capability of ETA model characterizing and predicting the iPSC culture regulation mechanism)}%This assumption explains the reason that most ETA prediction of metabolites at 12-h to 36-h is located between the predictions of metabolites at 24-h and 48-h obtained by using the developed metabolic kinetic model.}

{Under conditions of limited substrate availability, there are significant changes in consumption rates, as exemplified by glucose in LGLL and LGHL cultures, as well as pyruvate in all four conditions. Unlike averaging approaches (e.g., ETA model), %relying on the assumption of metabolic steady-state, 
the proposed metabolic kinetic model effectively captures these essential dynamic behaviors. This capability holds promise for facilitating future research on the end-to-end cell culture process optimal control to improve high-quality iPSC production. %Under limited substrate conditions, the consumption rate changes significantly, as seen with glucose on LGLL and LGHL, and pyruvate on all four conditions. In contrast to methodologies based on the metabolic steady-state assumption, the proposed metabolic kinetic model captures these critical dynamic behaviors, which can facilitate the development of bioprocess control implementation of fed-batch bioprocessing conditions for iPSCs is feasible in the future.
 Additionally, as stirred suspension cultures become a common choice for large-scale iPSC manufacturing \cite{borys2020optimized,meng2017optimizing}, this study can also contribute to the understanding of cell behavior within aggregates, as the well-established pattern involves low glucose and high lactate concentrations at the aggregate center \citep{kinney2011multiparametric}.}

{
By modeling the regulatory mechanisms outlined in Section~\ref{subsec:regulatorymechanism}, the results presented in Table~\ref{tab:pred_flux} are consistent with observations reported in existing studies \cite{ghorbaniaghdam2013kinetic, ghorbaniaghdam2014analyzing, ghorbaniaghdam2014silico, ivarsson2015insights, mulukutla2012metabolic, costa2007lactate, glacken1988mathematical, hassell1991growth, halestrap2012monocarboxylate, huckabee1956control}.
The ability of the proposed model to accurately predict cell dynamic behavior and response under varying environmental conditions reaffirms its reliability and underscores its potential to enhance the optimization and control of iPSC culture processes.
Specifically, the model indicates that the consumption rate of glucose increases with higher extracellular glucose concentrations (e.g., HGLL vs. LGLL and HGHL vs. LGHL), while it decreases with higher extracellular lactate concentrations (e.g., HGLL vs. HGHL and LGLL vs. LGHL). These trends can be attributed to regulatory mechanism \textbf{R1} as described in equation~(\ref{equ:R1}). Regarding lactate, the model reveals that the net production rate is lower when a certain amount of lactate exists in the cultural environment (e.g., HGLL vs. HGHL and LGLL vs. LGHL). In the case of pyruvate, under high lactate culture conditions, more favorable lactate transport kinetics of MCTs result in decreased pyruvate consumption (e.g., HGLL vs. HGHL and LGLL vs. LGHL). These observations are associated with regulatory mechanism \textbf{R6} as explained in equation~(\ref{equ:R6}). Furthermore, the presence of lactate dampens the conversion of glutamine to glutamate. Consequently, under high lactate culture conditions, the consumption rate of glutamine and the production rate of glutamate decrease (e.g., HGLL vs. HGHL and LGLL vs. LGHL). These behaviors are linked to regulatory mechanisms \textbf{R2} in equation~(\ref{equ:R2}) and \textbf{R7} in equation~(\ref{equ:R7}).}

\begin{table*}[h!]
\centering
\caption{{Dynamic model prediction for biomass-specific uptake and production rates for key extracellular metabolite (nmol/10$^6$ cells$\cdot$h). The fluxes obtained by  Extracellular Time-Course Analysis (ETA)  for 12-h  to 36-h are calculated averages from measurements \citep{odenwelder2021induced}, while the 12-h and 36-h fluxes predicted by the developed metabolic kinetic model are calculated for that specific time point.  Negative values represent consumption rates and positive values represent production rates.}}
\label{tab:pred_flux}
\begin{tabular}{|c|ccc|ccc|}
\hline
\rowcolor[HTML]{DDEBF7} 
& \multicolumn{3}{c|}{\cellcolor[HTML]{DDEBF7}\textbf{HGLL   (nmol/10$^6$ cells$\cdot$h)}}                                                                       & \multicolumn{3}{c|}{\cellcolor[HTML]{DDEBF7}\textbf{LGLL   (nmol/10$^6$ cells$\cdot$h)}}                                                                       \\ \hline
\rowcolor[HTML]{DDEBF7} 
\textbf{Metabolite}                                     & \multicolumn{1}{c|}{\cellcolor[HTML]{DDEBF7}\textbf{12-h}} & \multicolumn{1}{c|}{\cellcolor[HTML]{DDEBF7}\textbf{36-h}} & \textbf{12-h to 36-h (ETA)} & \multicolumn{1}{c|}{\cellcolor[HTML]{DDEBF7}\textbf{12-h}} & \multicolumn{1}{c|}{\cellcolor[HTML]{DDEBF7}\textbf{36-h}} & \textbf{12-h to 36-h (ETA)} \\ \hline
\cellcolor[HTML]{DDEBF7}\textbf{Glucose}                & \multicolumn{1}{c|}{-1,910}                               & \multicolumn{1}{c|}{-1,132}                                & -1,051                       & \multicolumn{1}{c|}{-1,664}                               & \multicolumn{1}{c|}{-826}                                & -994                        \\ \hline
\cellcolor[HTML]{DDEBF7}\textbf{Lactate}                & \multicolumn{1}{c|}{1,410}                                & \multicolumn{1}{c|}{2,145}                                & 2,108                        & \multicolumn{1}{c|}{1,400}                                & \multicolumn{1}{c|}{2,096}                                & 2,000                        \\ \hline
\cellcolor[HTML]{DDEBF7}\textbf{Pyruvate}               & \multicolumn{1}{c|}{-109}                                 & \multicolumn{1}{c|}{-58.2}                                 & -67.7                       & \multicolumn{1}{c|}{-109}                                 & \multicolumn{1}{c|}{-60.2}                                 & -75.2                       \\ \hline
\cellcolor[HTML]{DDEBF7}\textbf{Glutamate}              & \multicolumn{1}{c|}{35.4}                                  & \multicolumn{1}{c|}{53.3}                                  & 36.2                        & \multicolumn{1}{c|}{35.5}                                  & \multicolumn{1}{c|}{54.4}                                  & 32.4                        \\ \hline
\cellcolor[HTML]{DDEBF7}\textbf{Glutamine}              & \multicolumn{1}{c|}{-221}                                & \multicolumn{1}{c|}{-180}                                & -132                        & \multicolumn{1}{c|}{-224}                                & \multicolumn{1}{c|}{-178}                                & -147                        \\ \hline
\rowcolor[HTML]{DDEBF7} 
\multicolumn{1}{|l|}{\cellcolor[HTML]{DDEBF7}\textbf{}} & \multicolumn{3}{c|}{\cellcolor[HTML]{DDEBF7}\textbf{HGHL (nmol/10$^6$ cells$\cdot$h)}}                                                                         & \multicolumn{3}{c|}{\cellcolor[HTML]{DDEBF7}\textbf{LGHL (nmol/10$^6$ cells$\cdot$h)}}                                                                         \\ \hline
\rowcolor[HTML]{DDEBF7} 
\textbf{Metabolite}                                     & \multicolumn{1}{c|}{\cellcolor[HTML]{DDEBF7}\textbf{12-h}} & \multicolumn{1}{c|}{\cellcolor[HTML]{DDEBF7}\textbf{36-h}} & \textbf{12-h to 36-h (ETA)} & \multicolumn{1}{c|}{\cellcolor[HTML]{DDEBF7}\textbf{12-h}} & \multicolumn{1}{c|}{\cellcolor[HTML]{DDEBF7}\textbf{36-h}} & \textbf{12-h to 36-h (ETA)} \\ \hline
\cellcolor[HTML]{DDEBF7}\textbf{Glucose}                & \multicolumn{1}{c|}{-880}                                & \multicolumn{1}{c|}{-773}                                & -874                        & \multicolumn{1}{c|}{-763}                                & \multicolumn{1}{c|}{-628}                                & -860                        \\ \hline
\cellcolor[HTML]{DDEBF7}\textbf{Lactate}                & \multicolumn{1}{c|}{1,332}                                 & \multicolumn{1}{c|}{1,602}                                & 1,694                        & \multicolumn{1}{c|}{1,314}                                 & \multicolumn{1}{c|}{1,388}                                & 1,651                        \\ \hline
\cellcolor[HTML]{DDEBF7}\textbf{Pyruvate}               & \multicolumn{1}{c|}{-71.2}                                 & \multicolumn{1}{c|}{-53.4}                                 & -39                         & \multicolumn{1}{c|}{-71.9}                                 & \multicolumn{1}{c|}{-53.3}                                 & -36.5                       \\ \hline
\cellcolor[HTML]{DDEBF7}\textbf{Glutamate}              & \multicolumn{1}{c|}{31.1}                                  & \multicolumn{1}{c|}{49.2}                                  & 28.3                        & \multicolumn{1}{c|}{31.6}                                  & \multicolumn{1}{c|}{49.8}                                  & 29.4                        \\ \hline
\cellcolor[HTML]{DDEBF7}\textbf{Glutamine}              & \multicolumn{1}{c|}{-206}                                & \multicolumn{1}{c|}{-161}                                & -107                        & \multicolumn{1}{c|}{-202}                                & \multicolumn{1}{c|}{-160}                                & -120                        \\ \hline
\end{tabular}
\end{table*}

\section{Conclusions}
\label{sec:conclusion}

\begin{sloppypar}
{
In this paper, a metabolic kinetic model is described that characterizes the time-varying dynamics and regulatory mechanisms of the iPSC cultures. %In this paper, we create a metabolic kinetic model characterizing the time-variate dynamics and regulatory mechanisms of the iPSC culture process. 
This model detailed central carbon metabolism, including glycolysis, TCA cycle, PPP, anaplerosis, and key amino acid metabolism. The iPSC metabolic regulatory network was calibrated using extracellular metabolite concentrations and intracellular isotopic data for multiple tracers (i.e., [1,2-$^{13}$C$_2$] glucose and [U-$^{13}$C$_{5}$] glutamine tracers). %and [U-$^{13}$C$_3$] sodium L-lactate tracers). 
These time-course measurements were collected at multiple time points for the different culture conditions. %(i.e., different levels of extracellular glucose and lactate concentrations). 
%Both the capacities of iPSC culture process forward prediction under the same experimental initial condition and extrapolation across different initial conditions were assessed by using experimental measures. 
%The developed metabolic kinetic model shows promising results and it provides a reliable prediction of stem cellular metabolic response to extracellular glucose and lactate concentration’s perturbation. It can advance the understanding of the iPSC culture intracellular regulatory mechanisms, improve the capability for real-time cell culture process monitoring, and support optimal control for integrated iPSC culture processes. 
The validation results demonstrate that the developed metabolic kinetic model can provide %shows great promise by providing 
a reliable prediction on iPSC metabolic response to changes in extracellular glucose and lactate levels.}

{This model holds substantial potential for advancing the understanding of complex intracellular regulatory mechanisms within iPSC cultures. %Furthermore, 
%It supports \textit{in silico} experimentation  for cell cultures under different initial concentrations and guides process control strategies. 
Furthermore, it empowers continuous monitoring and precise control of cell cultures, particularly advantageous in the realm of large-scale manufacturing. 
%Due to strong cell-to-cell interactions, iPSCs form large cell aggregates resulting in insufficient nutrient supply and extra metabolic waste build-up for the cells located at the core. 
%In challenging scenarios such as stirred suspension cultures, where maintaining adequate nutrient levels (e.g., glucose) and managing metabolic waste (e.g., lactate) within aggregates are intricate tasks, This mechanistic model plays a pivotal role in comprehending the consequences of micro-environmental fluctuations. 
In summary, the proposed mechanistic model holds the potential to facilitate the optimization and control of iPSC cultures, potentially even at large scales, ensuring the survival, productivity, and quality of iPSC-derived cell products.}
%In challenging contexts like stirred suspension cultures, where ensuring nutrient availability (e.g., glucose) and managing metabolic waste (e.g., lactate) are challenging, this model can be invaluable in understanding the effects arising from environmental variations. }
\end{sloppypar}

%\newpage
%\bibliographystyle{wileyNJD-APA}
\bibliography{reference}

%\clearpage
\newpage
\appendix
%\begin{appendix}
\onecolumn

\section{Appendix: Table}
%In this section, 
%\section{Stoichiometry of Metabolic Reactions}
The following tables summarize detailed information about the constructed iPSC metabolic network and developed kinetic model.
\begin{itemize}
    \item The iPSC metabolic network contains the major reactions for glycolysis, the TCA cycle, anaplerosis, PPP, and amino acid metabolism; see the specific reactions in Table~\ref{tab:reaction}. The reactions of PPP are collapsed into two: Oxidative phase/branch and Non-oxidative phase/branch (i.e., No. 9 \& 10 reactions). 
    \item The iPSC metabolic flux rate regulation biokinetic model is summarized in Table~\ref{tab:kinetic}. The Michaelis-Menten model is used to characterize how key activators and inhibitors influence regulatory mechanisms in each reaction.
    \item The descriptions of the metabolites, considered in the developed metabolic kinetic model, are listed in Table~\ref{tab:metabolite}.
    \item For the developed metabolic kinetic model, the descriptions of the enzymes with Enzyme Commission Number (EC-No.) are provided in Table~\ref{tab:enzyme}. They are associated with metabolic flux reaction rates.

\end{itemize}

\clearpage
\begin{table}
\centering
\caption{Reactions of the metabolic network}
\label{tab:reaction}
\begin{tabular}{|l|l|}
\hline
\rowcolor[HTML]{C0C0C0} 
\textbf{No.} & \multicolumn{1}{c|}{\textbf{Pathway}}                    \\ \hline \hline
             & \multicolumn{1}{c|}{\textbf{Glycolysis}}   \\ \hline
\rowcolor[HTML]{C0C0C0} 
\textbf{1}   & Glc(abcdef) $\rightarrow$ G6P(abcdef)             \\ \hline
\textbf{2}   & G6P(abcdef)$\rightarrow$ F6P(abcdef)              \\ \hline
\rowcolor[HTML]{C0C0C0} 
\textbf{3}   & F6P(abcdef) $\rightarrow$ GAP(cba) + GAP (def)    \\ \hline
\textbf{4}   & GAP(abc) $\rightarrow$ PEP(abc)                   \\ \hline
\rowcolor[HTML]{C0C0C0} 
\textbf{5}   & PEP(abc) $\rightarrow$ Pyr(abc)                   \\ \hline
\textbf{6}   & Pyr(abc) $\leftrightarrow$ Lac(abc)                   \\ \hline
\rowcolor[HTML]{C0C0C0} 
\textbf{7}   & EPyr(abc) $\rightarrow$ Pyr(abc)                 \\ \hline
\textbf{8}   & Lac(abc) $\leftrightarrow$ ELac(abc)                 \\ \hline \hline
\rowcolor[HTML]{C0C0C0} 
             & \multicolumn{1}{c|}{\textbf{PPP}}   \\ \hline
\textbf{9}   & G6P(abcdef)$\rightarrow$ Ru5P(bcdef)+CO$_2$(a)       \\ \hline
\rowcolor[HTML]{C0C0C0} 
\textbf{10}  & Ru5P(abcde)+Ru5P(fghij)+Ru5P(klmno) $\rightarrow$ F6P(fgahij) +F6P(klbcde)+ GAP(mno)              \\ \hline \hline
             & \multicolumn{1}{c|}{\textbf{TCA}}    \\ \hline
\rowcolor[HTML]{C0C0C0} 
\textbf{11}  & Pyr(abc)$\rightarrow$ AcCoA(bc)+CO$_2$(a)            \\ \hline
\textbf{12}  & AcCoA(ab)+OAA(cdef)$\rightarrow$ Cit(fedbac)      \\ \hline
\rowcolor[HTML]{C0C0C0} 
\textbf{13}  & Cit(abcdef) $\leftrightarrow$ AKG(abcde) +CO$_2$(f)      \\ \hline
\textbf{14}  & AKG(abcde) $\rightarrow$ Suc(bcde)+ CO$_2$(a)        \\ \hline
\rowcolor[HTML]{C0C0C0} 
\textbf{15}  & Suc(abcd)$\rightarrow$ Fum(abcd)                  \\ \hline
\textbf{16}  & Fum(abcd)$\leftrightarrow$ Mal(abcd)                  \\ \hline
\rowcolor[HTML]{C0C0C0} 
\textbf{17}  & Mal(abcd)$\leftrightarrow$ OAA(abcd)                  \\ \hline \hline
             & \multicolumn{1}{c|}{\textbf{Anaplerosis and Amino Acid}}    \\ \hline
\rowcolor[HTML]{C0C0C0} 
\textbf{18}  & Mal(abcd) $\rightarrow$ Pyr(abc)+CO$_2$(d)           \\ \hline
\textbf{19}  & Pyr(abc) +CO$_2$(d) $\rightarrow$ OAA(abcd)          \\ \hline
\rowcolor[HTML]{C0C0C0} 
\textbf{20}  & Gln(abcde) $\leftrightarrow$ Glu(abcde)+NH$_4$           \\ \hline
\textbf{21}  & Glu(abcde)$\leftrightarrow$ AKG(abcde)+NH$_4$            \\ \hline
\rowcolor[HTML]{C0C0C0} 
\textbf{22}  & Glu(abcde)+Pyr(fgh)$\leftrightarrow$ AKG(abcde)+Ala(fgh)                     \\ \hline
\textbf{23}  & Ala(abc)$\rightarrow$ EAla(abc)                   \\ \hline
\rowcolor[HTML]{C0C0C0} 
\textbf{24}  & Glu(abcde)$\rightarrow$ EGlu(abcde)               \\ \hline
\textbf{25}  & EGln(abcde)$\rightarrow$ Gln(abcde)               \\ \hline
\rowcolor[HTML]{C0C0C0} 
\textbf{26}  & Ser(abc) $\rightarrow$ Pyr(abc)+ NH$_4$               \\ \hline
\textbf{27}  & Asp(fghi)+AKG(abcde) $\leftrightarrow$ Glu(abcde)+OAA(fghi)+NH$_4$               \\ \hline
\rowcolor[HTML]{C0C0C0} 
\textbf{28}  & EAsp(abcd) $\rightarrow$ Asp(abcd)                 \\ \hline
\textbf{29}  & Cit(abcdef) $\rightarrow$ Mal(fcba) + Lipids       \\ \hline \hline
\rowcolor[HTML]{C0C0C0} 
             & \multicolumn{1}{c|}{\textbf{Biomass}}                       \\ \hline
\textbf{30}  & 0.19 Ala + 0.11 Asp + 0.1 Gln + 0.12 Glu + 0.17 Gly + 0.14 Ser + 0.16 Glc $\rightarrow$ Biomass \\ \hline
\end{tabular}
\end{table}

%\newpage
\clearpage
\begin{table}[hbt!]
\centering
\caption{Biokinetic equations of the metabolites fluxes (1-30) of the model}
\label{tab:kinetic}
\begin{tabular}{|lp{13.5cm}|}
\hline
\rowcolor[HTML]{C0C0C0} 
\textbf{No.} & \multicolumn{1}{c|}{\textbf{Pathway}}                    \\ \hline \hline
             & \multicolumn{1}{c|}{\textbf{Glycolysis}}   \\ \hline
\rowcolor[HTML]{C0C0C0} 
\multicolumn{1}{|l|}{\textbf{1}}   & $v(HK)= v_{max, HK} \times \frac{Glc}{K_{m, Glc}+Glc} \times \frac{K_{i, G6P}}{K_{i, G6P}+G6P} \times \frac{K_{i, LactoHK}}{K_{i, LactoHK}+Lac}$    \\ \hline
\multicolumn{1}{|l|}{\textbf{2}}   & $v(PGI)= v_{max, PGI} \times \frac{G6P}{K_{m, G6P}+G6P}$   \\ \hline
\rowcolor[HTML]{C0C0C0} 
\multicolumn{1}{|l|}{\textbf{3}}   & $v(PFK/ALD) = v_{max, PFK/ALD} \times \frac{F6P}{K_{m, F6P}+F6P}$                \\ \hline
\multicolumn{1}{|l|}{\textbf{4}}   & $v(PGK) = v_{max, PGK} \times \frac{GAP}{K_{m, GAP}+GAP}$   \\ \hline
\rowcolor[HTML]{C0C0C0} 
\multicolumn{1}{|l|}{\textbf{5}}   & $v(PK) = v_{max, PK} \times \frac{PEP}{K_{m, PEP} \times (1+\frac{K_{a, F6P}}{F6P})+PEP}$          \\ \hline
\multicolumn{1}{|l|}{\textbf{6f}}    & $v{LDHf}= v_{max, fLDH} \times \frac{Pyr}{K_{m, Pyr}+Pyr}$                       \\ \hline
\rowcolor[HTML]{C0C0C0} 
\multicolumn{1}{|l|}{\textbf{6r}}   & $v(LDHr) = v_{max, rLDH} \times \frac{Lac}{K_{m, Lac}+Lac} \times \frac{K_{i, Pyr}}{K_{i, Pyr}+Pyr}$   \\ \hline
\multicolumn{1}{|l|}{\textbf{7}}   & $v(PyrT) = v_{max, PyrT} \times \frac{EPyr}{K_{m, EPyr}+EPyr} \times \frac{K_{i, LactoPyr}}{K_{i, LactoPyr}+Lac}$              \\ \hline
\rowcolor[HTML]{C0C0C0} 
\multicolumn{1}{|l|}{\textbf{8f}}   & $v(LacTf) = v_{max, fLacT} \times \frac{Lac}{K_{m, Lac}+Lac}$                     \\ \hline
\multicolumn{1}{|l|}{\textbf{8r}}    & $v(LacTr) = v_{max, rLacT} \times \frac{ELac}{K_{m, ELac}+ELac}$                  \\ \hline  \hline
\rowcolor[HTML]{C0C0C0} 
\multicolumn{1}{|l|}{\textbf{}}    & \multicolumn{1}{c|}{\textbf{PPP}}                   \\ \hline
\multicolumn{1}{|l|}{\textbf{9}}   & $v(OP) = v_{max, OP} \times \frac{G6P}{K_{m, G6P}+G6P}$   \\ \hline
\rowcolor[HTML]{C0C0C0} 
\multicolumn{1}{|l|}{\textbf{10}}   & $v(NOP) = v_{max, NOP} \times \frac{Ru5P}{K_{m, Ru5P}+Ru5P}$                     \\ \hline  \hline
\multicolumn{1}{|l|}{\textbf{}}    & \multicolumn{1}{c|}{\textbf{TCA}}                   \\ \hline
\rowcolor[HTML]{C0C0C0} 
\multicolumn{1}{|l|}{\textbf{11}}   & $v(PDH) = v_{max, PDH} \times \frac{Pyr}{K_{m, Pyr}+Pyr}$   \\ \hline
\multicolumn{1}{|l|}{\textbf{12}}    & $v(CS) = v_{max, CS} \times \frac{AcCoA}{K_{m, AcCoA}+AcCoA} \times \frac{OAA}{K_{m, OAA}+OAA}$      \\ \hline
\rowcolor[HTML]{C0C0C0} 
\multicolumn{1}{|l|}{\textbf{13f}}   & $v(CITS/ISODf) = v_{max, fCITS/ISOD} \times \frac{Cit}{K_{m, Cit}+Cit}$          \\ \hline
\multicolumn{1}{|l|}{\textbf{13r}}   & $v(CITS/ISODr) = v_{max, rCITS/ISOD} \times \frac{AKG}{K_{m, AKG}+AKG}$          \\ \hline
\rowcolor[HTML]{C0C0C0} 
\multicolumn{1}{|l|}{\textbf{14}}   & $v(AKGDH) = v_{max, AKGDH} \times \frac{AKG}{K_{m, AKG}+AKG}$                     \\ \hline
\multicolumn{1}{|l|}{\textbf{15}}    & $v(SDH) = v_{max, SDH} \times \frac{Suc}{K_{m, SUC}+Suc}$                        \\ \hline
\rowcolor[HTML]{C0C0C0} 
\multicolumn{1}{|l|}{\textbf{16f}}                        & $v(FUMf) = v_{max, fFUM} \times \frac{Fum}{K_{m, Fum}+Fum}$                      \\ \hline
\multicolumn{1}{|l|}{\textbf{16r}}   & $v(FUMr) = v_{max, rFUM} \times \frac{Mal}{K_{m, Mal}+Mal}$                     \\ \hline
\rowcolor[HTML]{C0C0C0} 
\multicolumn{1}{|l|}{\textbf{17f}}                        & $v(MDHf) = v_{max, fMDH} \times \frac{Mal}{K_{m, Mal}+Mal}$                       \\ \hline
\multicolumn{1}{|l|}{\textbf{17r}}   & $v(MDHr) = v_{max, rMDH} \times \frac{OAA}{K_{m, OAA}+OAA}$                       \\ \hline
\end{tabular}
\end{table}

%~\newpage
%\section{Biokinetic Model of Flux Rate Regulation Mechanism}
%The iPSC metabolic flux rate regulation biokinetic model is summarized in Table~\ref{tab:kinetic}. Key activators and inhibitors are considered for each reaction, and their impact on the regulatory mechanisms is characterized through Michaelis–Menten model.

\begin{table}[t!]
\centering
%\caption{Biokinetic equations of the metabolites fluxes (1-30) of the model}
%\label{tab:kinetic}
\begin{tabular}{|lp{13.5cm}|}
\hline
\rowcolor[HTML]{C0C0C0} 
\multicolumn{1}{|l|}{\textbf{}}    & \multicolumn{1}{c|}{\textbf{Anaplerosis and Amino Acid}}                   \\ \hline
\multicolumn{1}{|l|}{\textbf{18}}    & $v(ME) = v_{max, ME} \times \frac{Mal}{K_{m, Mal}+Mal}$                        \\ \hline
\rowcolor[HTML]{C0C0C0} 
\multicolumn{1}{|l|}{\textbf{19}}   & $v(PC) = v_{max, PC} \times \frac{Pyr}{K_{m, Pyr}+Pyr}$                        \\ \hline
\multicolumn{1}{|l|}{\textbf{20f}}   & $v(GLNSf) = v_{max, fGLNS} \times \frac{Gln}{K_{m, Gln}+Gln} \times \frac{K_{i, LactoGLNS}}{K_{i, LactoGLNS}+Lac}$             \\ \hline
\rowcolor[HTML]{C0C0C0} 
\multicolumn{1}{|l|}{\textbf{20r}}                        & $v(GLNSr) = v_{max, rGLNS} \times \frac{Glu}{K_{m, Glu}+Glu} \times \frac{NH_4}{K_{m, NH_4}+NH_4}$      \\ \hline
\multicolumn{1}{|l|}{\textbf{21f}}   & $v(GLDHf) = v_{max, fGLDH} \times \frac{Glu}{K_{m, Glu}+Glu}$                     \\ \hline
\rowcolor[HTML]{C0C0C0} 
\multicolumn{1}{|l|}{\textbf{21r}}                        & $v(GLDHr) = v_{max, rGLDH} \times \frac{AKG}{K_{m, AKG}+AKG} \times \frac{NH_4}{K_{m, NH_4}+NH_4}$      \\ \hline
\multicolumn{1}{|l|}{\textbf{22f}}   & $v(AlaTAf) = v_{max, fAlaTA} \times \frac{Glu}{K_{m, GLU}+Glu} \times \frac{Pyr}{K_{m, Pyr}+Pyr}$    \\ \hline
\rowcolor[HTML]{C0C0C0} 
\multicolumn{1}{|l|}{\textbf{22r}}                        & $v(AlaTAr) = v_{max, rAlaTA} \times \frac{Ala}{K_{m, Ala}+Ala} \times \frac{AKG}{K_{m, AKG}+AKG} \times (1+\frac{K_{a, Gln}}{Gln})$             \\ \hline
\multicolumn{1}{|l|}{\textbf{23}}    & $v(AlaT) = v_{max, AlaT} \times \frac{Ala}{K_{m, Ala}+Ala}$                       \\ \hline
\rowcolor[HTML]{C0C0C0} 
\multicolumn{1}{|l|}{\textbf{24}}   & $v(GluT) = v_{max, GluT} \times \frac{Glu}{K_{m, Glu}+Glu}$                       \\ \hline
\multicolumn{1}{|l|}{\textbf{25}}    & $v(GlnT) = v_{max, GlnT} \times \frac{EGln}{K_{m, EGln}+EGln} \times \frac{K_{i, GLN}}{K_{i, GLN}+GLN}$                        \\ \hline
\rowcolor[HTML]{C0C0C0} 
\multicolumn{1}{|l|}{\textbf{26}}   & $v(SAL) = v_{max, SAL} \times \frac{Ser}{K_{m, Ser}+Ser}$                \\ \hline
\multicolumn{1}{|l|}{\textbf{27f}}   & $v(ASTAf) = v_{max, fASTA} \times \frac{Asp}{K_{m, ASP}+Asp} \times \frac{AKG}{K_{m, AKG}+AKG}$         \\ \hline
\rowcolor[HTML]{C0C0C0} 
\multicolumn{1}{|l|}{\textbf{27r}}   & $v(ASTAr) = v_{max, rASTA} \times \frac{Glu}{K_{m, Glu}+Glu} \times \frac{OAA}{K_{m, OAA}+OAA} \times \frac{NH_4}{K_{m, NH_4}+NH_4}$                 \\ \hline
\multicolumn{1}{|l|}{\textbf{28}}    & $v(AspT) = v_{max, AspT} \times \frac{EAsp}{K_{m, EAsp}+EAsp}$                       \\ \hline
\rowcolor[HTML]{C0C0C0} 
\multicolumn{1}{|l|}{\textbf{29}}  & $v(ACL) = v_{max, ACL} \times \frac{Cit}{K_{m, Cit}+Cit}$    \\ \hline  \hline
\multicolumn{1}{|l|}{}             & \multicolumn{1}{c|}{\textbf{Biomass}}               \\ \hline
\rowcolor[HTML]{C0C0C0} 
\multicolumn{1}{|l|}{\textbf{30}}  & $v(growth) = v_{max, growth} \times \frac{Gln}{K_{m, Gln}+Gln} \times \frac{Glc}{K_{m, Glc}+Glc} \times \frac{Glu}{K_{m, Glu}+Glu} \times \frac{Ala}{K_{m, Ala}+Ala} \times \frac{Asp}{K_{m, Asp}+Asp} \times \frac{Ser}{K_{m, Ser}+Ser} \times \frac{Gly}{K_{m, Gly}+Gly}$ \\ \hline
\end{tabular}
\end{table}

\clearpage
%\section{Abbreviation of Metabolites}
%The descriptions of the metabolites, considered in the developed metabolic kinetic model, are listed in Table~\ref{tab:metabolite}.

\begin{table}[h!]
\centering
\caption{Description of the Metabolite}
\label{tab:metabolite}
\begin{tabular}{|l|l|l|l|}
\hline
\textbf{Component} & \textbf{Description}       & \textbf{Component} & \textbf{Description}    \\ \hline \hline
\rowcolor[HTML]{B7B7B7} 
ACCoA              & Acetyl-CoezymeA            & ALA                & Alanine                 \\ \hline
AKG                & $\alpha$-Ketoglutarate            & ASP                & Aspartate               \\ \hline
\rowcolor[HTML]{B7B7B7} 
CIT                & Citrate                    & LAC                & Lactate                 \\ \hline
CO2                & Intracellular Carbonoxygen & GLN                & Glutamine               \\ \hline
\rowcolor[HTML]{B7B7B7} 
F6P                & Fructose 6-Phosphate       & EGLY               & Extracellular Glycine   \\ \hline
G6P                & Glucose 6-Phosphate        & SER                & Extracellular Serine    \\ \hline
\rowcolor[HTML]{B7B7B7} 
GAP                & Glyceraldehyde 3-Phosphate & GLC                & Extracellular Glucose   \\ \hline
GLU                & Glutamate                  & EGLN               & Extracellular Glutamine \\ \hline
\rowcolor[HTML]{B7B7B7} 
GLY                & Glycine                    & EGLU               & Extracellular Glutamate \\ \hline
MAL                & Malate                     & EPYR               & Extracellular Pyruvate  \\ \hline
\rowcolor[HTML]{B7B7B7} 
OAA                & Oxaloacetate               & EASP               & Extracellular Aspartate \\ \hline
PEP                & Phosphoenolpyruvate        & EALA               & Extracellular Alanine   \\ \hline
\rowcolor[HTML]{B7B7B7} 
FUM                & Fumarate                   & ELAC               & Extracellular Lactate   \\ \hline
Ru5P               & Ribulose 5-Phosphate       & NH4                & Extracellular Ammonia   \\ \hline
\rowcolor[HTML]{B7B7B7} 
SUC                & Succinate                  & LIPID              & Lipid                   \\ \hline
PYR                & Pyruvate                   & Bio                & Biomass                 \\ \hline
\end{tabular}
\end{table}

\clearpage
%\newpage
%\section{Abbreviation of Enzymes}
%For the developed metabolic kinetic model, the descriptions of the enzymes with Enzyme Commission Number (EC-No.) are provided in Table~\ref{tab:enzyme}. They are associated with metabolic flux reaction rates.

\begin{table}[!bh!]
\centering
\caption{Description of the Enzyme}
\label{tab:enzyme}
\begin{tabular}{|l|l|l|}
\hline
\rowcolor[HTML]{FFFFFF} 
\textbf{Abbreviation} & \textbf{Description}               & \textbf{EC-No.}   \\ \hline  \hline
\rowcolor[HTML]{C0C0C0} 
HK                    & Hexokinase                         & 2.7.1.1           \\ \hline
\rowcolor[HTML]{FFFFFF} 
PGI                   & Phosphoglucose Isomerase           & 5.3.1.9           \\ \hline
\rowcolor[HTML]{C0C0C0} 
PFK/ALD               & Phosphofructokinase/Aldolase       & 2.7.1.11/4.1.2.13 \\ \hline
PGK                   & Phosphoglycerate Kinase            & 2.7.2.3           \\ \hline
\rowcolor[HTML]{C0C0C0} 
PK                    & Pyruvate Kinase                    & 2.7.1.40          \\ \hline
OP                    & Oxidative Phase of PPP             &                   \\ \hline
\rowcolor[HTML]{C0C0C0} 
NOP                   & Non-oxidative Phase of PPP         &                   \\ \hline
PyrT                  & Membrane Transport of Pyruvate     &                   \\ \hline
\rowcolor[HTML]{C0C0C0} 
SAL                   & Membrane Transport of Serine       &                   \\ \hline
LDH                   & Lactate Dehydrogenase              & 1.1.1.27          \\ \hline
\rowcolor[HTML]{C0C0C0} 
AlaTA                 & Alanine Transaminase               & 2.6.1.2           \\ \hline
PC                    & Pyruvate Carboxylase               & 6.4.1.1           \\ \hline
\rowcolor[HTML]{C0C0C0} 
PDH                   & Pyruvate Dehydrogenase             & 1.2.4.1           \\ \hline
CS                    & Citrate (Si)-Synthase              & 2.3.3.1           \\ \hline
\rowcolor[HTML]{C0C0C0} 
CITS/ISOD             & Aconitase/Isocitrate Dehydrogenase & 4.2.1.3/1.1.1.41  \\ \hline
GLDH                  & Glutamate Dehydrogenase            & 1.4.1.2           \\ \hline
\rowcolor[HTML]{C0C0C0} 
GluT                  & Membrane Transport of Glutamate    &                   \\ \hline
GLNS                  & Glutamine Synthetase               & 6.3.1.2           \\ \hline
\rowcolor[HTML]{C0C0C0} 
AKGDH                 & $\alpha$-ketoglutarate Dehydrogenase      & 1.2.1.105         \\ \hline
SDH                   & Succinate Dehydrogenase            & 1.3.5.1           \\ \hline
\rowcolor[HTML]{C0C0C0} 
MDH                   & Malate Dehydrigenase               & 1.1.1.37          \\ \hline
ME                    & Malic Enzyme                       & 1.1.1.40          \\ \hline
\rowcolor[HTML]{C0C0C0} 
ASTA                  & Aspartate Aminotransferase         & 2.6.1.1           \\ \hline
ACL                   & ATP citrate synthase               & 2.3.3.8           \\ \hline
\rowcolor[HTML]{C0C0C0} 
FUM                   & Fumarase                           & 4.2.1.2           \\ \hline
\end{tabular}
\end{table}

\clearpage
\section{Appendix: Cell Characteristic Predictions}
\begin{figure*}[h!]
    \centering
    \includegraphics[width=1\textwidth]{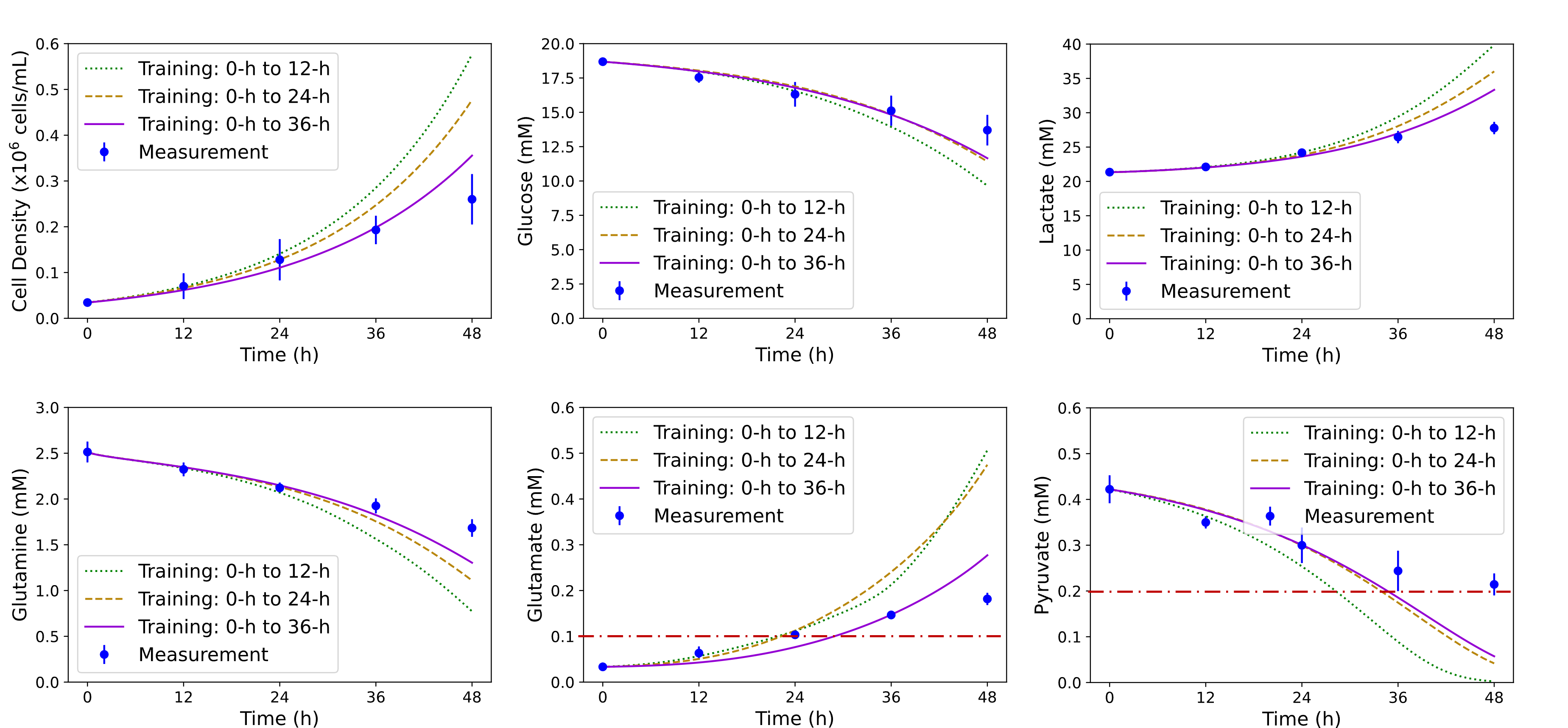}
    \caption{{Dynamic model trained with different time intervals - Cell characteristic predictions for high glucose and high lactate cultures. (A) Cell density, (B) Glucose, (C) Lactate, (D) Glutamine, (E) Glutamate, and (F) Pyruvate. Times 0-h to 12-h (green dotted line); Times 0-h to 24-h (brown dashed line); and Times 0-h to 36-h (purple solid line). The detection limit of the Cedex Bioanalyzer is shown as the red dash-dot line.  }}%{\textbf{Various Time Interval:} Cell characteristic predictions for the high glucose and high lactate cultures using the dynamic model trained on different time intervals. (A) Cell density, (B) Glucose, (C) Lactate, (D) Glutamine, (E) Glutamate, and (F) Pyruvate. Times 0-h to 12-h (green dotted line); Times 0-h to 24-h (brown dashed line); and Times 0-h to 36-h (purple solid line). The detection limit of the Cedex Bioanalyzer is shown as the red dash-dot line.}} 
    \label{fig:pred_forward_hghl}
\end{figure*}

\begin{figure*}[h!]
    \centering
    \includegraphics[width=1\textwidth]{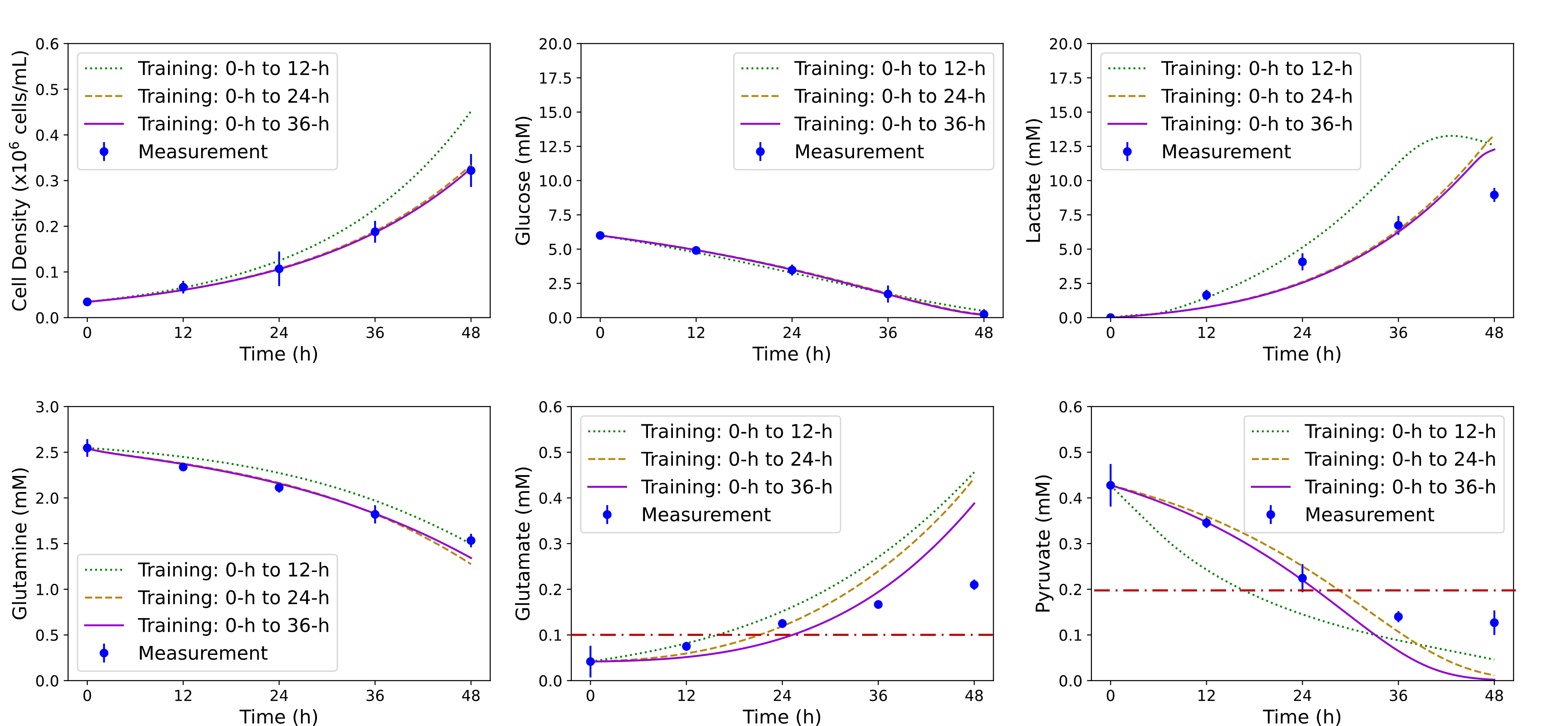}
    \caption{{Dynamic model trained with different time intervals - Cell characteristic predictions for low glucose and low lactate cultures. (A) Cell density, (B) Glucose, (C) Lactate, (D) Glutamine, (E) Glutamate, and (F) Pyruvate. Times 0-h to 12-h (green dotted line); Times 0-h to 24-h (brown dashed line); and Times 0-h to 36-h (purple solid line). The detection limit of the Cedex Bioanalyzer is shown as the red dash-dot line.  }}%{\textbf{Various Time Interval:} Cell characteristic predictions for the low glucose and low lactate cultures using the dynamic model trained on different time intervals. (A) Cell density, (B) Glucose, (C) Lactate, (D) Glutamine, (E) Glutamate, and (F) Pyruvate. Times 0-h to 12-h (green dotted line); Times 0-h to 24-h (brown dashed line); and Times 0-h to 36-h (purple solid line). The detection limit of the Cedex Bioanalyzer is shown as the red dash-dot line.}} 
    \label{fig:pred_forward_lgll}
\end{figure*}

\begin{figure*}[h!]
    \centering
    \includegraphics[width=1\textwidth]{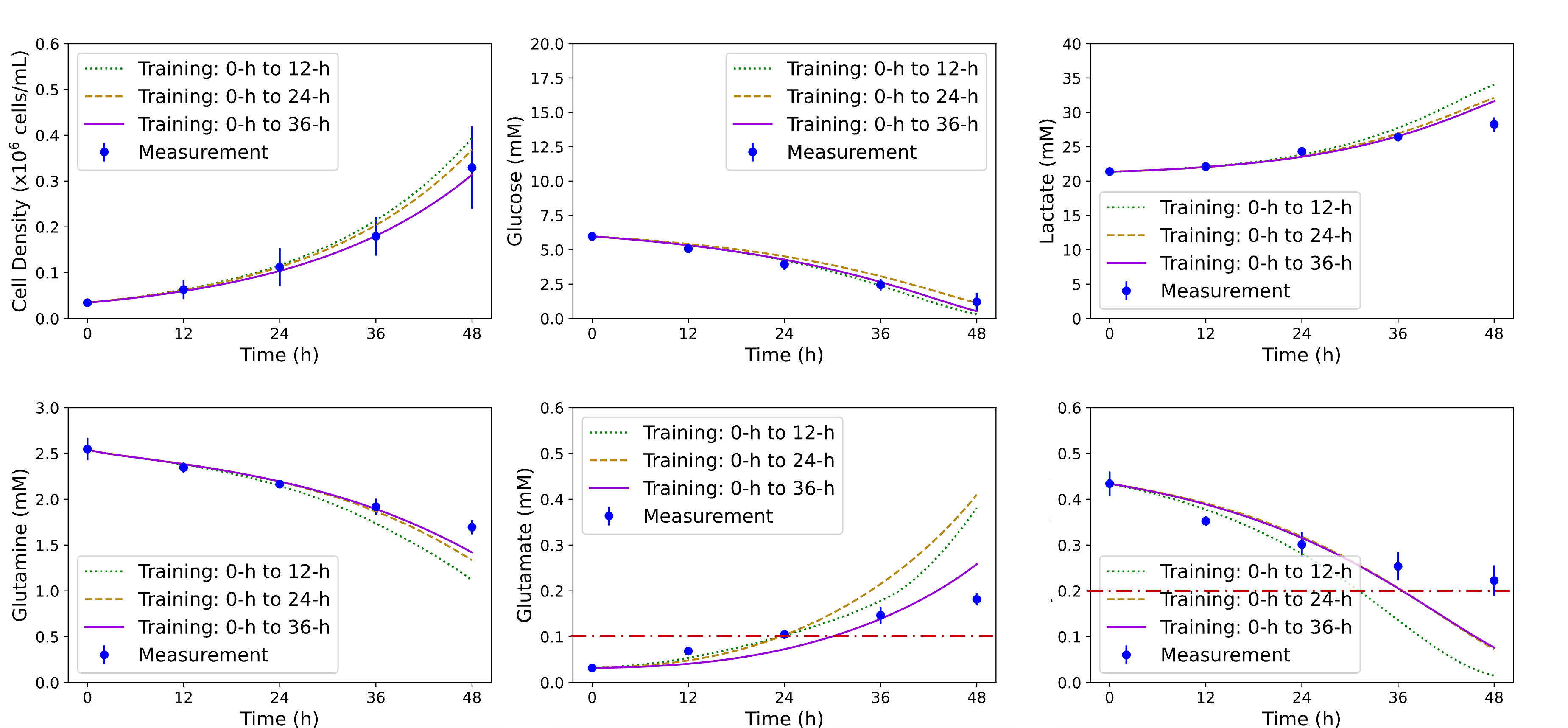}
    \caption{{Dynamic model trained with different time intervals - Cell characteristic predictions for low glucose and high lactate cultures. (A) Cell density, (B) Glucose, (C) Lactate, (D) Glutamine, (E) Glutamate, and (F) Pyruvate. Times 0-h to 12-h (green dotted line); Times 0-h to 24-h (brown dashed line); and Times 0-h to 36-h (purple solid line). The detection limit of the Cedex Bioanalyzer is shown as the red dash-dot line.  }}%{\textbf{Various Time Interval:} Cell characteristic predictions for the low glucose and high lactate cultures using the dynamic model trained on different time intervals. (A) Cell density, (B) Glucose, (C) Lactate, (D) Glutamine, (E) Glutamate, and (F) Pyruvate. Times 0-h to 12-h (green dotted line); Times 0-h to 24-h (brown dashed line); and Times 0-h to 36-h (purple solid line). The detection limit of the Cedex Bioanalyzer is shown as the red dash-dot line.}} 
    \label{fig:pred_forward_lghl}
\end{figure*}

{Figures \ref{fig:pred_forward_hghl} to \ref{fig:pred_forward_lghl} depict the model's predictions for iPSC cultures in three scenarios: high glucose and high lactate (HGHL), low glucose and low lactate (LGLL), and low glucose and high lactate cultures (LGHL). Predictions are based on training the model using data collected over various time intervals.}

\clearpage
\section{Appendix: Cell Characteristic Across Culture Condition Predictions}
\begin{figure*}[h!]
    \centering
    \includegraphics[width=1\textwidth]{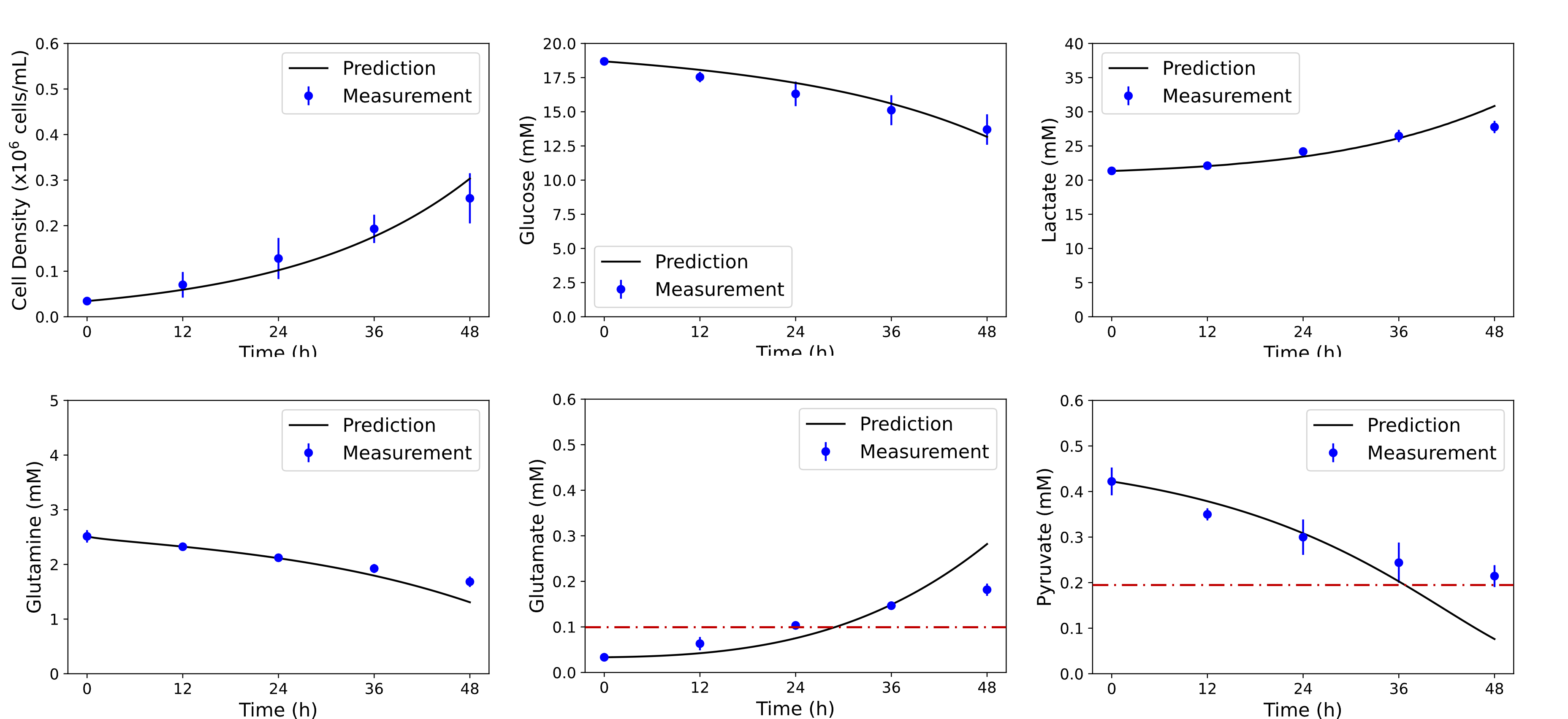}
    \caption{{Dynamic model trained with three other datasets with varied initial glucose and lactate concentrations for cell characteristic predictions in high glucose and high lactate cultures.  (A) Cell density, (B) Glucose, (C) Lactate, (D) Glutamine, (E) Glutamate, and (F) Pyruvate. The detection limit of the Cedex Bioanalyzer is shown as the red dash-dot line.}} %{\textbf{Different Initial Conditions:} Cell characteristic predictions for the high glucose and high lactate cultures using the dynamic model trained on the other three case data sets.  (A) Cell density, (B) Glucose, (C) Lactate, (D) Glutamine, (E) Glutamate, and (F) Pyruvate.  The detection limit of the Cedex Bioanalyzer is shown as the red dash-dot line.}} 
    \label{fig:pred_across_hghl}
\end{figure*}

\begin{figure*}[h!]
    \centering
    \includegraphics[width=1\textwidth]{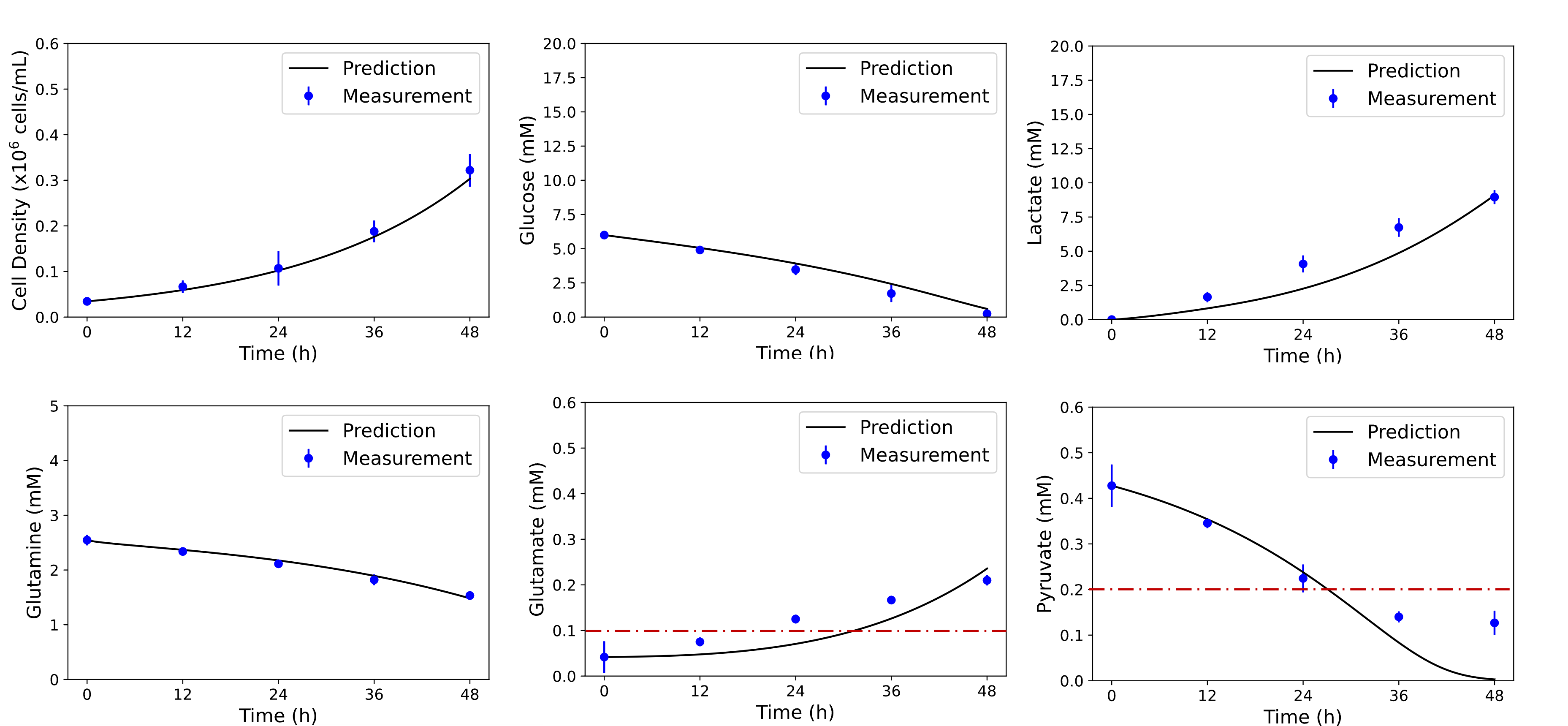}
    \caption{{Dynamic model trained with three other datasets with varied initial glucose and lactate concentrations for cell characteristic predictions in low glucose and low lactate cultures.  (A) Cell density, (B) Glucose, (C) Lactate, (D) Glutamine, (E) Glutamate, and (F) Pyruvate. The detection limit of the Cedex Bioanalyzer is shown as the red dash-dot line.}}%{\textbf{Different Initial Conditions:} Cell characteristic predictions for the low glucose and low lactate cultures using the  dynamic model trained on the other three case data sets.  (A) Cell density, (B) Glucose, (C) Lactate, (D) Glutamine, (E) Glutamate, and (F) Pyruvate.  The detection limit of the Cedex Bioanalyzer is shown as the red dash-dot line.}} 
    \label{fig:pred_across_lgll}
\end{figure*}

\begin{figure*}[h!]
    \centering
    \includegraphics[width=1\textwidth]{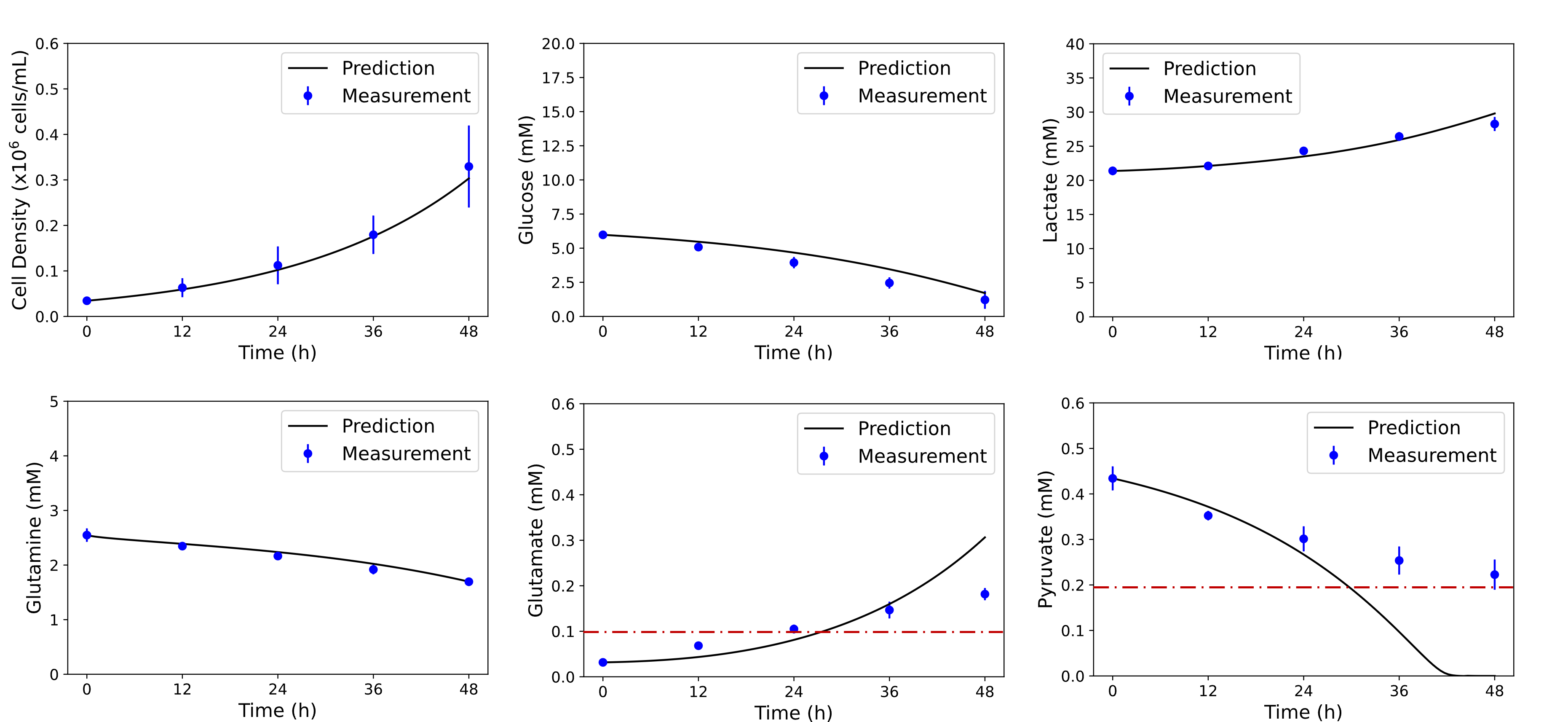}
    \caption{{Dynamic model trained with three other datasets with varied initial glucose and lactate concentrations for cell characteristic predictions in low glucose and high lactate cultures.  (A) Cell density, (B) Glucose, (C) Lactate, (D) Glutamine, (E) Glutamate, and (F) Pyruvate. The detection limit of the Cedex Bioanalyzer is shown as the red dash-dot line.}}%{\textbf{Different Initial Conditions:} Cell characteristic predictions for the low glucose and high lactate cultures using the  dynamic model trained on the other three case data sets.  (A) Cell density, (B) Glucose, (C) Lactate, (D) Glutamine, (E) Glutamate, and (F) Pyruvate.  The detection limit of the Cedex Bioanalyzer is shown as the red dash-dot line.}} 
    \label{fig:pred_across_lghl}
\end{figure*}

{Figures \ref{fig:pred_across_hghl} to \ref{fig:pred_across_lghl} depict prediction results across various initial conditions, including high glucose and high lactate (HGHL), low glucose and low lactate (LGLL), and low glucose and high lactate (LGHL). The model's fitting is based on data from the remaining three cases.} 

\clearpage
\section{Appendix: Metabolic flux maps}
\begin{figure}[h!]%
    \centering
    \subfloat{{\includegraphics[width=0.47\textwidth]{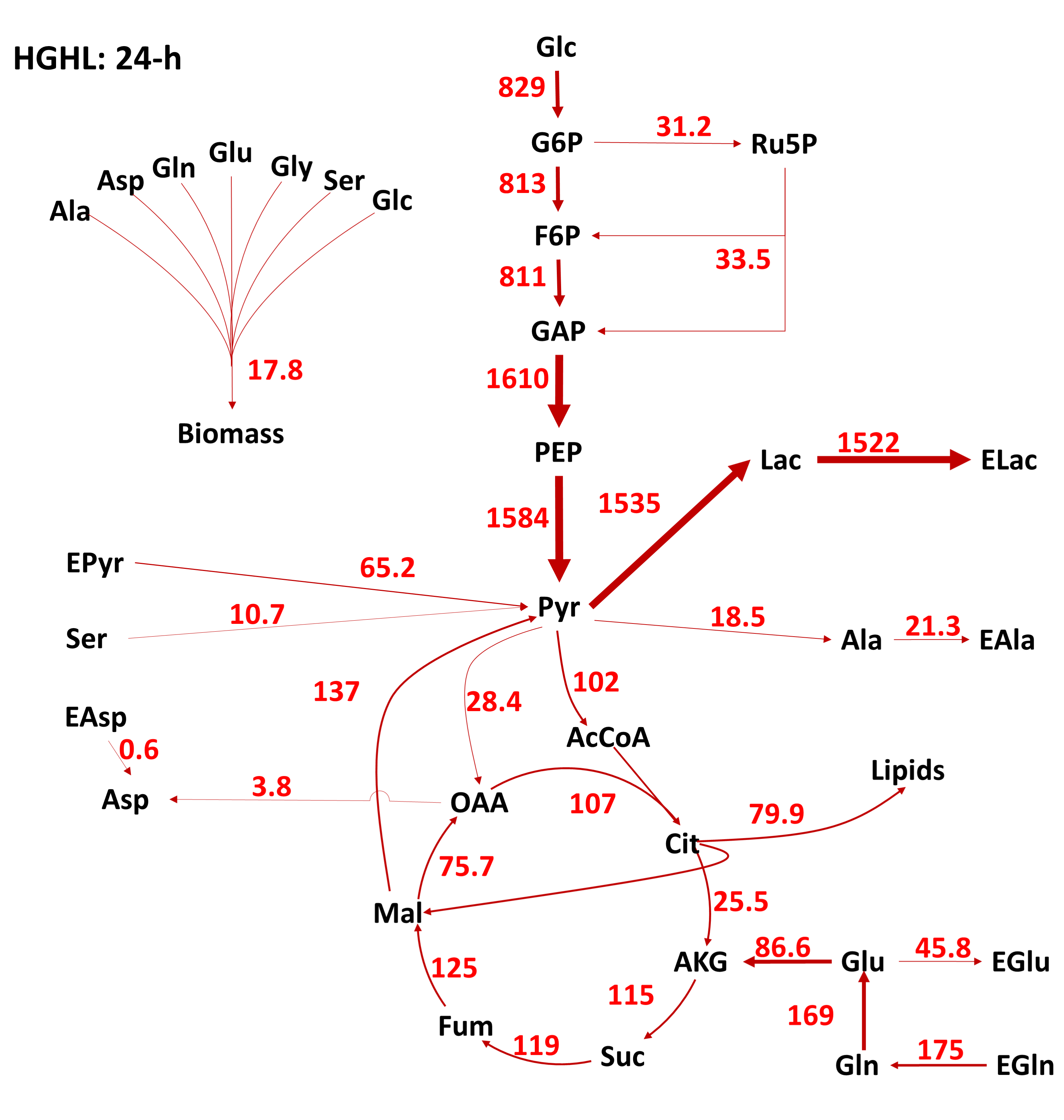} }}%
    \qquad
    \subfloat{{\includegraphics[width=0.47\textwidth]{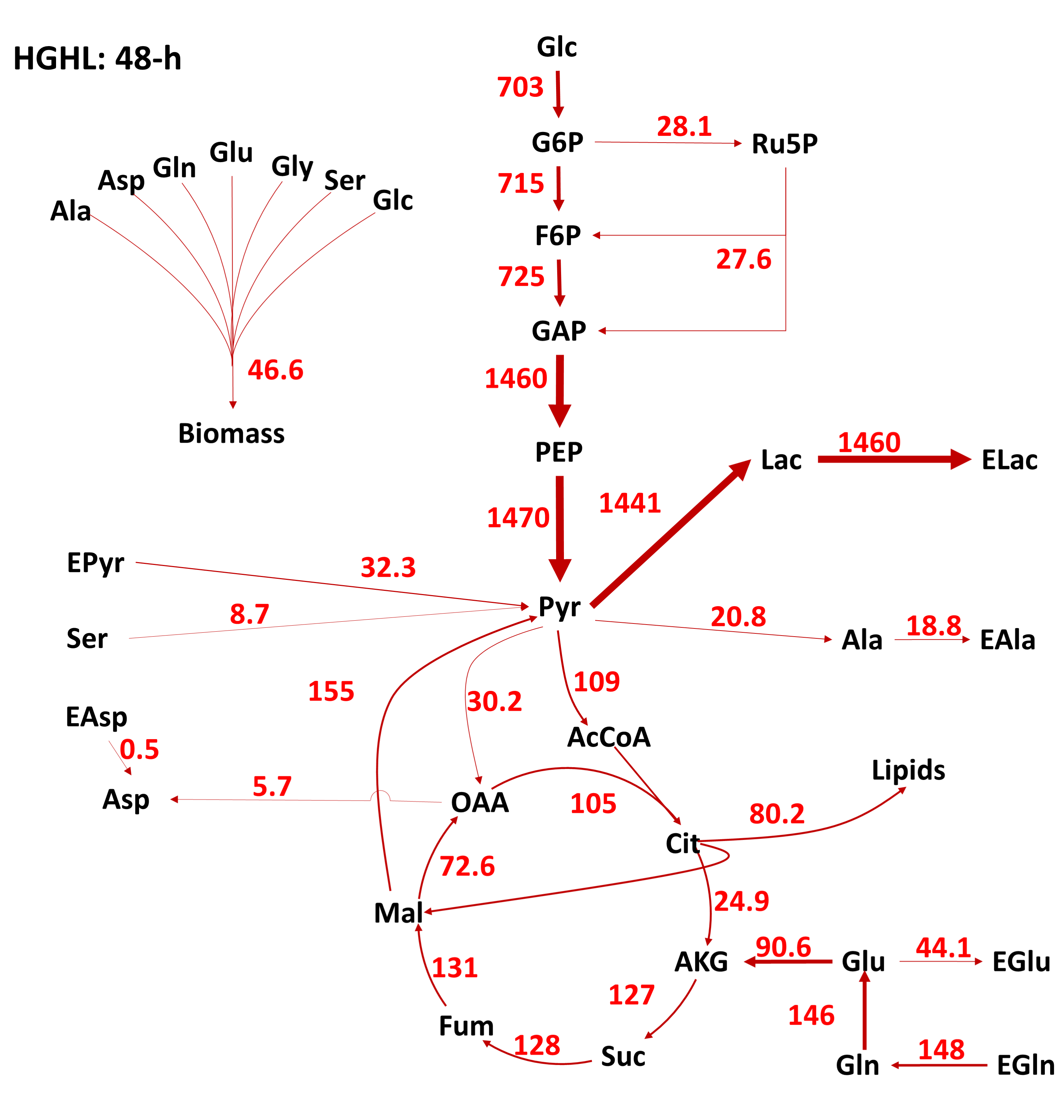} }}%
    \caption{Metabolic flux maps for K3 iPSC for the high glucose and high lactate cultures at 24-h and 48-h.  Predicted fluxes are given in nmol/10$^6$ cells$\cdot$h. The line thicknesses represent the relative fluxes.}%
    \label{fig:fluxmap_HGHL}%
\end{figure}

\begin{figure*}[h!]%
    \centering
    \subfloat{{\includegraphics[width=0.47\textwidth]{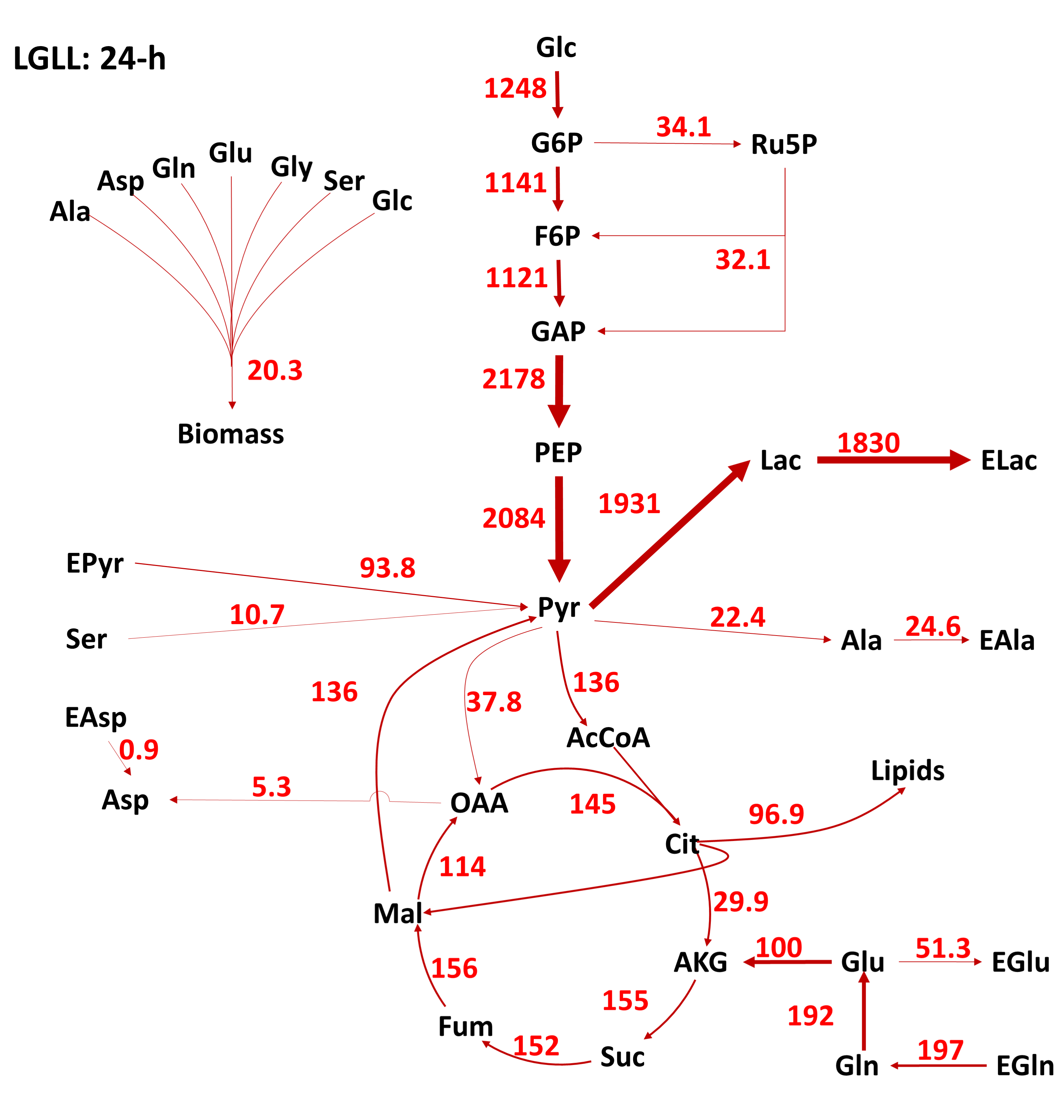} }}%
    \qquad
    \subfloat{{\includegraphics[width=0.47\textwidth]{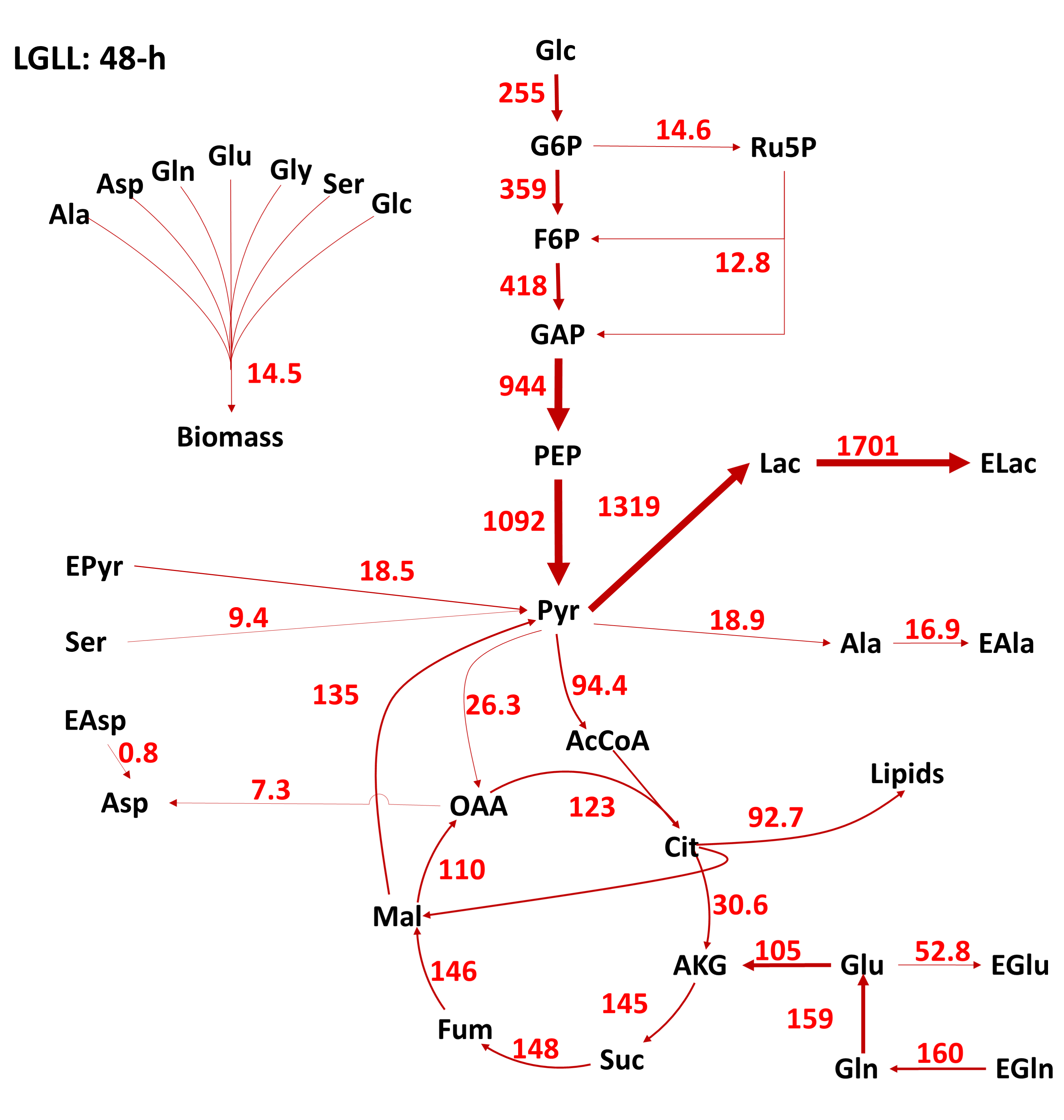} }}%
    \caption{Metabolic flux maps for K3 iPSC for the low glucose and low lactate cultures at 24-h and 48-h.  Predicted fluxes are given in nmol/10$^6$ cells$\cdot$h. The line thicknesses represent the relative fluxes.}%
    \label{fig:fluxmap_LGLL}%
\end{figure*}

\begin{figure*}[h!]%
    \centering
    \subfloat{{\includegraphics[width=0.47\textwidth]{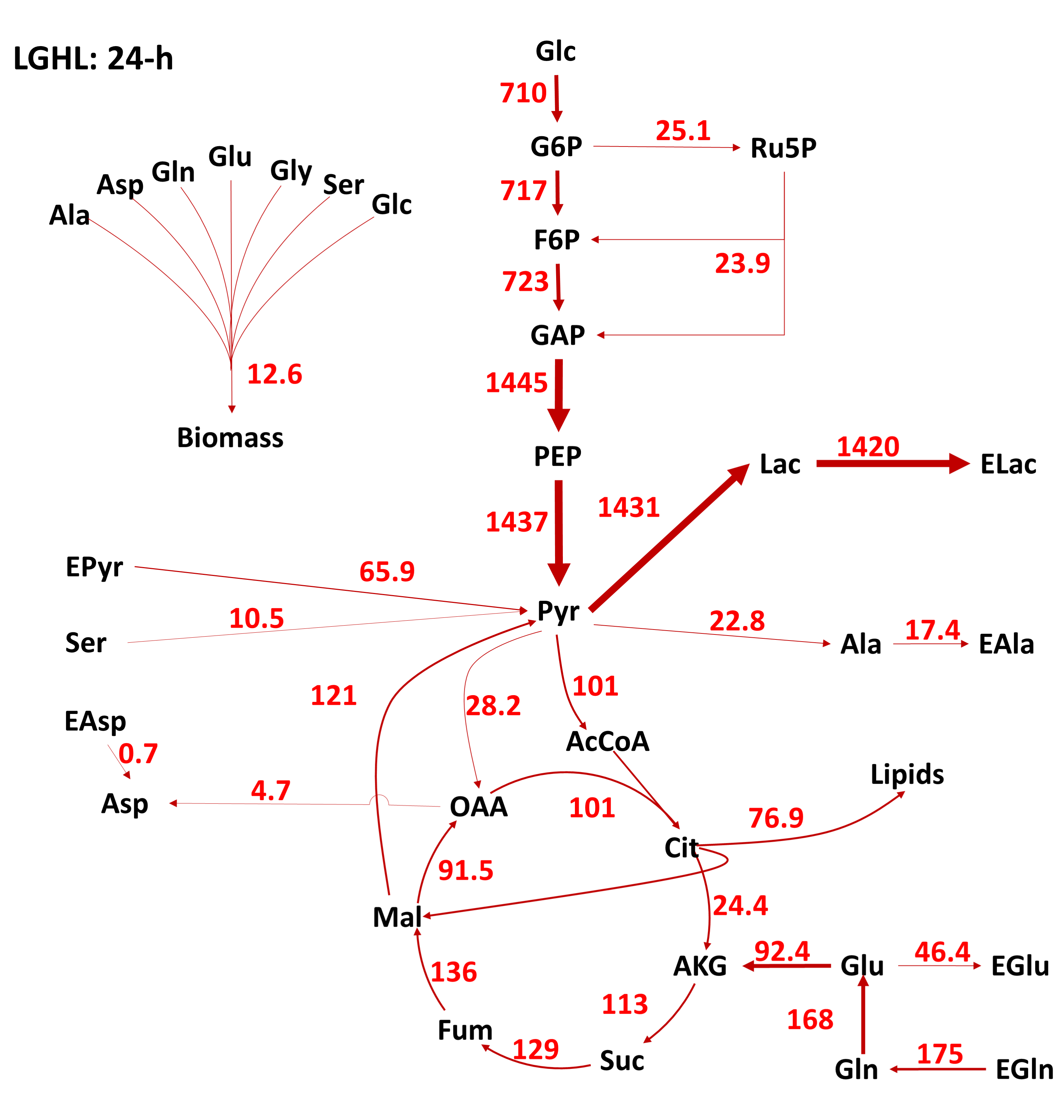} }}%
    \qquad
    \subfloat{{\includegraphics[width=0.47\textwidth]{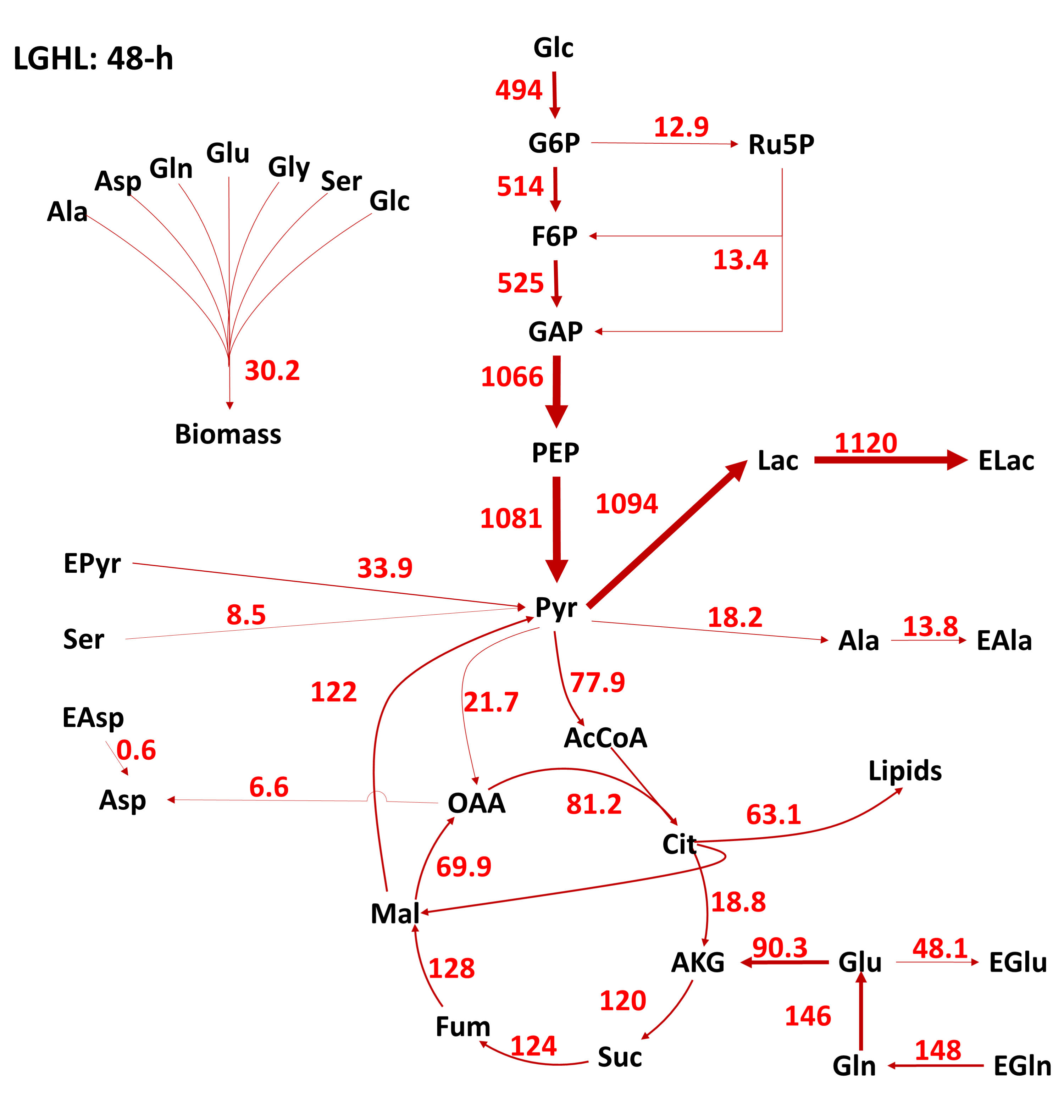} }}%
    \caption{Metabolic flux maps for K3 iPSC for the low glucose and high lactate cultures at 24-h and 48-h.  Predicted fluxes are given in nmol/10$^6$ cells$\cdot$h. The line thicknesses represent the relative fluxes.}%
    \label{fig:fluxmap_LGHL}%
\end{figure*}

Fig~\ref{fig:fluxmap_HGHL} to \ref{fig:fluxmap_LGHL} depict the metabolic flux maps for iPSC culture at 24-h and 48-h respectively under the initial conditions, including high glucose and high lactate (HGHL), low glucose and low lactate (LGLL), and low glucose and high lactate (LGHL).  

\newpage
%\end{appendix}
\end{document}